\documentstyle[preprint,revtex]{aps}

\begin{document}
 
\draft
\begin{title}
{\large \bf Antiresonance and Localization in Quantum Dynamics}
\end{title}
 
\author{\bf I. Dana, E. Eisenberg, and N. Shnerb}
\begin{instit}
Department of Physics, Bar-Ilan University, Ramat-Gan 52900, Israel
\end{instit}
 
\begin{abstract}
The  phenomenon of  quantum antiresonance  (QAR), i.e.,  exactly periodic
recurrences  in  quantum  dynamics,  is  studied  in  a  large  class  of
nonintegrable systems, the  modulated kicked rotors (MKRs).   It is shown
that asymptotic  exponential localization generally occurs  for $\eta$ (a
scaled $\hbar$)  in the  infinitesimal vicinity  of QAR  points $\eta_0$.
The  localization  length  $\xi_0$  is  determined  from  the  analytical
properties  of  the  kicking   potential.   This  ``QAR-localization"  is
associated in  some cases with  an integrable limit of  the corresponding
classical systems.  The MKR dynamical problem is mapped into pseudorandom
tight-binding  models,   exhibiting  dynamical  localization   (DL).   By
considering  exactly-solvable cases,  numerical  evidence  is given  that
QAR-localization is  an excellent approximation to  DL sufficiently close
to QAR.   The transition from  QAR-localization to DL in  a semiclassical
regime,  as  $\eta$  is  varied,  is studied.   It  is  shown  that  this
transition takes  place via a gradual  reduction of the influence  of the
analyticity of  the potential on  the analyticity of the  eigenstates, as
the level of chaos is increased.
\end{abstract}
 
 
\narrowtext
 
\newpage
 
\begin{center}{\bf I. INTRODUCTION}\\ \end{center}
 
The study  of ``quantum chaos",  i.e., understanding the  fingerprints of
classical  chaos in  quantum  mechanics \cite{book,qc},  has  led to  the
discovery of a variety of  new quantum-dynamical phenomena.  Several such
phenomena  occur  in  time-periodic  systems  described  by  the  general
Hamiltonian
\begin{equation}\label{H}
H = H_0 + H_1 f(t)\ ,
\end{equation}
where   $H_0$  is   some   time-independent  Hamiltonian,   $H_1$  is   a
perturbation, and $f(t)$ is periodic  with period $T$, $f(t+T)=f(t)$.  In
many  cases,  $f(t)$ is  chosen,  for  simplicity,  as a  periodic  delta
function,   $f(t)=\Delta_T(t)\equiv    \sum_{s=-\infty}^{\infty}   \delta
(t-sT)$,   giving   the   well-known    class   of   ``kicked"   systems.
Representative   models   in   this    class   are   the   kicked   rotor
\cite{rot,izr,hog,fish,she,blu,fis,kr2,kr3,alt}, the  kicked Harper model
\cite{khm}, and the kicked harmonic oscillator \cite{kho}.\\
 
The quantum  dynamics of time-periodic  systems (\ref{H}) is  governed by
their quasienergy  (QE) spectrum  (i.e., the  spectrum of  the one-period
evolution operator).   Different properties  of the  QE spectrum  lead to
quantum-dynamical phenomena having, in general, no classical analogue.  A
classic example  is the quantum  suppression of chaotic diffusion  in the
kicked rotor  (KR) \cite{rot},  accompanied by  quasiperiodic recurrences
\cite{hog}.   An important  interpretation  of this  phenomenon has  been
given  \cite{fish,she} by  showing  first that,  in the  angular-momentum
representation,  the  QE  eigenstates  of the  KR  satisfy  the  equation
describing  a 1D  tight-binding model  with pseudorandom  disorder.  This
pseudorandomness is  generic, as  it exists  for almost  all (irrational)
values  of a  scaled (dimensionless)  $\hbar$,  which we  denote here  by
$\eta$.   It  was found  \cite{gri}  that  in several  interesting  cases
pseudorandom tight-binding models exhibit localization properties similar
to  those  of  truly  random  ones  (Anderson  localization)  \cite{mot}.
Assuming the  general occurrence of  this localization, it  follows that,
generically, the  QE eigenstates  are exponentially localized  in angular
momentum  and  the QE  spectrum  is  pure  point.  This  localization  in
pseudorandom  tight-binding   models,  equivalent   to  quantum-dynamical
systems  with  nonintegrable  classical   counterparts,  is  called  {\it
dynamical  localization} (DL)  \cite{not}.   The  quantum suppression  of
diffusion in the KR is an  immediate consequence of DL.  Despite the fact
that DL has  no classical analogue, there exists a  remarkable and simple
relation between  the classical chaotic-diffusion coefficient  $D$ in the
KR and the  {\it asymptotic} DL length $\xi$ in  the semiclassical regime
(sufficiently  small $\eta$):  $\xi \approx  D/2$ \cite{rot,she,blu,fis}.
For nongeneric, rational values of $\eta$, there occurs in all the kicked
systems  the  phenomenon  of  quantum  resonance  \cite{izr},  i.e.,  the
quadratic  increase  of the  energy  expectation-value  with time.   This
phenomenon is due  to an absolutely continuous QE  spectrum, exhibiting a
band structure.\\
 
In this  paper, DL  is approached  in the  light of  a different  kind of
phenomenon  for systems  (\ref{H}): exactly  {\it periodic}  recurrences.
This phenomenon is defined, in general, by
\begin{equation}\label{QAR}
U^p = e^{-i\beta}\ ,
\end{equation}
where   $U$  is   the  one-period   evolution  operator   for  (\ref{H}),
$e^{-i\beta}$ is  some constant phase factor  (a $c$ number), and  $p$ is
the smallest positive  integer for which (\ref{QAR})  is satisfied.  Thus
$pT$ is the recurrence period.  As it will become apparent in this paper,
the phenomenon (\ref{QAR})  may occur, in general, only  for very special
values of  $\eta$, and it is  thus nongeneric.  In fact,  for the general
class of systems introduced in this  paper, it occurs precisely at values
of $\eta =\eta_0$  corresponding to quantum resonances  \cite{izr} in the
kicked systems.  At the same time, this phenomenon, manifesting itself in
bounded,  periodic  variation  of expectation  values,  is  diametrically
opposite to quantum  resonance.  We shall therefore  refer to (\ref{QAR})
as to the {\it quantum antiresonance} (QAR) phenomenon.\\
 
While this  phenomenon is nongeneric, we  show in this paper  that, for a
large class  of nonintegrable systems,  it is generally accompanied  by a
very interesting effect: In the  immediate vicinity of QAR (infinitesimal
$\eta   -\eta_0$),  there   takes  place   {\it  asymptotic   exponential
localization}  with a  pure-point  QE spectrum.   The  existence of  this
``QAR-localization"  is  rigorously established  in  the  framework of  a
self-consistent approach, which allows for  an exact determination of the
asymptotic   localization   length    $\xi_0$.    For   $\eta_0=0$,   the
QAR-localization is associated with an  integrable limit of the classical
Hamiltonian.   On  the  other  hand, values  of  $\eta_0\neq  0$  usually
correspond to the  strong quantum regime of  a nonintegrable Hamiltonian,
exhibiting chaotic diffusion.\\
 
As in the  case of the KR and  other systems \cite{fish,she,blu,kr2,kr3},
we show that our class of systems can be mapped into tight-binding models
with pseudorandom  disorder (for $\eta$ sufficiently  irrational).  DL is
then  expected  to   occur  generically  in  our   systems.   For  $\eta$
infinitesimally close to $\eta_0$, the tight-binding models are {\it not}
defined,  so  that  QAR-localization  {\it cannot}  be  viewed,  strictly
speaking, as a kind of DL.  However, if $\eta$ is sufficiently irrational
and  close to  $\eta_0$, one  expects  DL to  take place  and to  exhibit
approximately  the  same  features  as  those  of  QAR-localization.   In
particular,  the DL  length  $\xi$  should be  well  approximated by  the
QAR-localization length  $\xi_0$, and  one expects that  $\xi \rightarrow
\xi_0$  as  $\eta  \rightarrow  \eta_0$.   We  provide  strong  numerical
evidence that this  is indeed the case.  Since $\xi_0$  can be determined
exactly, this seems to  be the first case where a DL  length $\xi$ can be
found with arbitrary accuracy in nonintegrable systems.\\
 
We also  study the dependence of  $\xi$ on $\eta -\eta_0$  for $\eta$ not
very close  to $\eta_0$.   This allows one  to understand  the transition
from QAR-localization to DL in regimes basically different in nature from
QAR.  Values of  $\eta$ sufficiently far from $\eta_0$  may correspond to
semiclassical regimes  of local or  global chaos.  As  already mentioned,
the semiclassical regime  of global chaos in KR  systems is characterized
by  the approximate  relation $\xi  \approx D/2$  \cite{rot,she,blu,fis}.
Calculations  of  $\xi$  for  the KR  were  performed  \cite{she,blu}  by
applying  the  method  of  minimal Lyapunov  exponent  \cite{lic}  to  an
equivalent pseudorandom  tight-binding model.  The  method is based  on a
finite  truncation   of  the  (generally  infinite)   vector  of  hopping
constants, so  that only  the first  $N$ neighbors  are kept.   Then $\xi
=\lim_{N\rightarrow \infty} 1/\gamma_N$, where  $\gamma_N$ is the minimal
Lyapunov exponent  of a  $2N$-dimensional symplectic map  associated with
the truncated model.  We show that  the truncated model has, for all $N$,
a well-defined  dynamical equivalent  exhibiting QAR.  Then,  by studying
numerically the dependence of $\gamma_N$ on  both $N$ and $\eta$, we show
that the transition from QAR-localization to DL in a semiclassical regime
($\xi \approx D/2$) takes place via  a gradual reduction of the influence
of the analyticity  of the system on the analyticity  of the eigenstates,
as the level of chaos is increased.\\
 
In  the  simple case  of  $N=1$,  i.e., a  nearest-neighbor  pseudorandom
``Lloyd model",  we derive the  {\it exact} relation  $\xi_0=D/2$.  Since
many numerical calculations indicate that  $\xi$ is independent of $\eta$
for  such a  model  (see,  e.g., Refs.  \cite{she,gri}),  this is  strong
evidence that the relation $\xi =D/2$ holds exactly for the corresponding
dynamical system.   A lengthy  derivation of this  relation was  given in
Ref. \cite{fis}, based  on the assumption that  the pseudorandom disorder
can be replaced by  a truly random one (this gives  the usual Lloyd model
\cite{llo}).\\
 
The paper  is organized as follows.   In Sec.  II, we  discuss some basic
aspects of  the QE  spectrum at  QAR, and consider  special cases  of QAR
occurring in ordinary KR systems.  In Sec.  III, we introduce the general
class of modulated kicked rotors  (MKRs), and determine values of $\eta$,
$\eta_0$, where  QAR of period  $p=1$ occurs for {\it  arbitrary} kicking
potentials [more  general cases of QAR,  of periods $p=1$ and  $p=2$, are
considered in Appendix A; in Appendix B, we study QAR of arbitrary period
for  integrable versions  of  the MKRs].   We  show, on  the  basis of  a
self-consistent   approach,  that   for   infinitesimal  $\eta   -\eta_0$
asymptotic  exponential localization  takes place,  with a  pure-point QE
spectrum.   This  spectrum and  the  QE  states  are determined  from  an
effective  Hamiltonian with  a periodic  potential, and  the localization
length $\xi_0$ is  fixed by the analytical properties  of this potential.
If $\eta_0=0$,  the effective  Hamiltonian turns out  to be  precisely an
integrable limit ($T\rightarrow 0$) of the classical MKR Hamiltonian.  In
Sec.  IV, we consider cases for which exact and closed results concerning
QAR-localization  (e.g.,  $\xi_0$)  and  associated QE  spectrum  can  be
obtained.  Using these results, we provide strong numerical evidence that
QAR-localization is  an excellent approximation to  DL sufficiently close
to QAR.  In Sec.  V, we show how MKR dynamical problems can be mapped, in
general,   into  multi-channel   tight-binding  models   \cite{der}  with
pseudorandom  disorder.   In  Sec.   VI, we  study  the  transition  from
QAR-localization to DL in a semiclassical  regime for a simple MKR system
equivalent to the KR.  This study is performed by considering the minimal
Lyapunov  exponents   $\gamma_N$  for   successive  truncations   of  the
corresponding  pseudorandom  tight-binding  model.   The  exact  relation
$\xi_0=D/2$ is  derived for a nearest-neighbor  pseudorandom Lloyd model.
Conclusions are  presented in Sec.  VII.   Some of our results  have been
briefly reported in Refs. \cite{us,des}.\\
 
 
\begin{center}  {\bf II.   QE SPECTRUM  AT QAR.   EXAMPLES OF  QAR IN  KR
SYSTEMS}\\ \end{center}
 
An  immediate consequence  of (\ref{QAR})  is  that the  spectrum of  $U$
consists  precisely of  $p$ eigenvalues  $\exp (-i\omega_l)$,  $l=0,...,\
p-1$, where the quasienergies $\omega_l$ are given by
\begin{equation}\label{QE}
\omega_l = \frac{\beta+2\pi l}{p} \ .
\end{equation}
Since  the QE  spectrum is  finite, each  quasienergy (\ref{QE})  must be
infinitely degenerate.  An  infinite set of QE states  associated with QE
level $l$ is  obtained by applying the  corresponding projection operator
for the  cyclic group $\{  e^{is\beta /p}U^s\}_{s=0,...,p-1}$ to  all the
states $\Psi$ in the Hilbert space:
\begin{equation}\label{QES}
\psi_l = \frac{1}{p}\sum_{s=0}^{p-1}e^{is\omega_l}U^s \Psi \ .
\end{equation}
We recall  here that  in the  case of quantum  resonance the  QE spectrum
consists of  a finite number  of bands  \cite{izr}.  The finite  width of
each of these bands leads to  ballistic motion (quadratic increase of the
energy expectation-value with time).  In the QAR case, on the other hand,
one has  the diametrically  opposite phenomenon of  periodic recurrences.
This phenomenon  has nothing to  do with  localized QE states,  since the
infinite basis of  states (\ref{QES}) for QE level $l$  can be chosen, of
course,  either localized  or  extended by  properly  choosing the  state
$\Psi$.  The periodic recurrences may be explained by saying that each of
the $p$ infinitely degenerate levels in the QAR case is the extreme limit
case of a quantum-resonance band of  zero width.  This point of view will
become clearer by the following examples.\\
 
A first case  of QAR was noticed by Izrailev  and Shepelyansky \cite{izr}
in the KR.  Consider the general KR Hamiltonian
\begin{equation}\label{KR}
H = \frac{L^2}{2I} + {\hat k}V(\theta )\Delta_T(t)\ ,
\end{equation}
where $L$ is  the angular momentum, $I$ is the  moment of inertia, ${\hat
k}$ is  a parameter, and $V(\theta  )$ is a general  periodic function of
the angle $\theta$.   The evolution operator for  (\ref{KR}), from $t=-0$
to $t=T-0$, is
\begin{equation}\label{UKR}
U = e^{-i\tau {\hat n}^2}e^{-ikV(\theta )}\ ,
\end{equation}
where ${\hat  n}\equiv L/\hbar  =-id/d\theta$, $\tau \equiv  \hbar T/2I$,
and $k\equiv {\hat k}/\hbar$.  Quantum  resonances occur, in general, for
rational values of $\eta =\tau /2\pi$ \cite{izr}.  Consider, however, the
special case of $\eta =1/2$.  Using the relation
\begin{equation}\label{pi}
e^{-i\pi {\hat n}^2} e^{-ikV(\theta )} =
e^{-ikV(\theta+\pi )} e^{-i\pi {\hat n}^2}\ ,
\end{equation}
which  is  easily established  by  comparing  the Fourier  expansions  of
$V(\theta )$ and $V(\theta +\pi )$, one finds in this case that
\[
U^2 = \exp \{ -ik[V(\theta )+V(\theta +\pi )]\} \ .
\]
Thus  the  condition (\ref{QAR})  for  QAR  is  satisfied with  $p=2$  if
$V(\theta  )+V(\theta +\pi  )=\beta  /k$ identically,  for some  $\beta$.
This implies that $V(\theta )$ must have the general Fourier expansion
\begin{equation}\label{fou}
V(\theta ) = \frac{\beta}{2k} +
\sum_{s=-\infty}^{\infty} v_{2s+1} e^{i(2s+1)\theta }\ .
\end{equation}
This is, of course, the case  for the standard potential $V(\theta )=\cos
(\theta )$, considered in Ref.  \cite{izr}.  According to (\ref{QE}), the
QE spectrum consists of two  infinitely degenerate levels, $\omega =\beta
/2,\ \beta /2+\pi$.  By ``switching on" even-harmonic components $v_{2s}$
in (\ref{fou}),  the infinite degeneracy  is removed, and the  two levels
broaden  into two  bands, corresponding  to the  generic spectrum  of the
$1/2$ quantum resonance.\\
 
More  general  results can  be  obtained  for  the ``linear"  version  of
(\ref{KR}) \cite{exa}, which is, however, integrable \cite{berry}:
\begin{equation}\label{LKR}
H = \frac{\tau}{T}L + {\hat k}V(\theta )\Delta _T(t)\ ,
\end{equation}
where $\tau$ is now some dimensionless parameter.  The evolution operator
for (\ref{LKR}) is
\begin{equation}\label{ULKR}
U = e^{-i\tau {\hat n}}e^{-ikV(\theta )}\ .
\end{equation}
The $p$th  power of  $U$ in  (\ref{ULKR}) can be  easily given  in closed
form:
\begin{equation}\label{ULKRp}
U^p = \exp \left [-ik\sum_{s=1}^p V(\theta -s\tau )\right ]
\exp (-ip\tau {\hat n})\ .
\end{equation}
Eq. (\ref{QAR})  is now satisfied  if and only if  $\sum_{s=1}^p V(\theta
-s\tau )  = \beta /k$ and  $\eta =\tau /2\pi =m/p$,  for relatively prime
integers  $m$  and  $p$.   The  latter  condition  (rational  $\eta$)  is
precisely  the general  condition for  quantum resonance  \cite{izr,exa}.
The former condition  gives, however, the opposite  phenomenon, i.e., the
QAR.  It  is easy  to see that  this condition is  satisfied only  if the
Fourier coefficients $v_n$ of $V(\theta )$ satisfy
\[
v_0 = \frac{\beta}{pk},\ \ \ \ \ \ v_{sp}=0\ \ \ (s\neq 0)\ .
\]
In  fact, $v_{sp}$  ($s\neq 0$)  are precisely  the Fourier  coefficients
which contribute to the  width of a QE band in the  case of $m/p$ quantum
resonance \cite{exa}.\\
 
These examples  show that in  ordinary KR systems  QAR may occur  only if
$V(\theta )$ satisfies some restrictive  conditions.  In the next section
we  shall introduce  systems  in  which QAR  occurs  for {\it  arbitrary}
kicking potentials, at some $\eta =\eta_0$.\\
 
 
\begin{center} {\bf  III.  QAR IN  MODULATED KR SYSTEMS  AND\\ ASYMPTOTIC
EXPONENTIAL LOCALIZATION}\\ \end{center}
 
We define the general modulated kicked rotor (MKR) by the Hamiltonian
\begin{equation}\label{MKR}
H = \frac{L^2}{2I} +
{\hat k}V(\theta )\sum_{j=0}^{M-1}c_j\Delta_T(t-t_j)\ ,
\end{equation}
where  $V(\theta  )$  is  an  analytic function  of  $\theta$,  and,  for
$j=0,...,\ M-1$, $c_j$ are arbitrary coefficients and
\[
0\leq t_j < t_{j+1} \leq T \ ,\ \ \ \ t_0=0\ ,\ \ t_M=T\ .
\]
The Hamiltonian  (\ref{MKR}) has the  general form (\ref{H})  with $f(t)=
C(t)\sum_{j=0}^{M-1}\Delta_T(t-t_j)$, where $C(t)$ is a periodic function
with  period  $T$,  satisfying  the $M$  conditions  $C(t_j)=c_j$.   Thus
(\ref{MKR}) may be viewed as a generalized KR with $M$ kicks at arbitrary
times  $t_j$ within  the  basic  period, and  modulated  by the  function
$C(t)$.  The classical map for (\ref{MKR}) is given by
\begin{equation}\label{map}
\begin{array}{rcl}
L_{s+1}      & = & L_s - {\hat k}c_jV'(\theta_s)\ , \\
\theta_{s+1} & = & \theta_s + [(t_{s+1}-t_s)/I] L_{s+1} \ ,
\end{array}
\end{equation}
where the integer $s$ is uniquely decomposed as $s=rM+j$ ($r$ integer and
$j=0,...,\ M-1$),  $t_s=rMT+t_j$, $L_s=L(t=t_s-0)$,  and $\theta_s=\theta
(t=t_s-0)$.   In  general, the  system  (\ref{MKR})  with (\ref{map})  is
classically  nonintegrable, and  exhibits  the transition  from local  to
global chaos  when ${\hat k}$  is increased, as  in the ordinary  KR case
\cite{diff};  see an  example in  Fig.\ \ref{logl}.   The simple  case of
$M=2$, with $c_0=-c_1=1$ and $t_1=T/2$  (the ``two-sided" KR) was studied
in detail in Refs. \cite{us,des}.  This case may already be considered as
an  approximation of  sinusoidal driving  potentials corresponding  to ac
electromagnetic  fields  \cite{gy}.    Better  approximations  should  be
achieved  by  using the  Hamiltonian  (\ref{MKR}),  with properly  chosen
coefficients  $c_j$.   The  study  in Refs.  \cite{us,des}  will  now  be
extended to the general case of (\ref{MKR}).\\
 
The evolution operator for (\ref{MKR}), from $t=-0$ to $t=T-0$, is
\begin{equation}\label{UMKR}
U = \prod_{j=0}^{M-1}\exp (-i\tau_j{\hat n}^2)\exp [-ic_jkV(\theta )]\ ,
\end{equation}
where, for $j=0,...,\ M-1$,
\begin{equation}\label{tauj}
\tau_j = \frac{\hbar (t_{j+1}-t_j)}{2I}\ ,
\end{equation}
and the factors under the product  sign in (\ref{UMKR}) are arranged from
right to left  in order of increasing $j$.  Now,  when all the quantities
(\ref{tauj}) are  integer multiples  of $2\pi$, i.e.,  $\tau_j=2\pi m_j$,
one  has $\exp  (-i\tau_j{\hat  n}^2)  = 1$  identically.   Then, if  the
coefficients $c_j$ satisfy the condition
\begin{equation}\label{s0}
\sum_{j=0}^{M-1} c_j = 0\ ,
\end{equation}
we find  that $U=1$ in  (\ref{UMKR}), corresponding  to a simple  case of
fundamental QAR ($p=1$).  For simplicity,  we shall restrict ourselves in
what follows to  this case, characterized by  the conditions $\tau_j=2\pi
m_j$  and (\ref{s0}).   More general  cases, for  both periods  $p=1$ and
$p=2$, are  considered in Appendix  A, where  it is conjectured  that QAR
with period $p>2$ do not exist for the MKR Hamiltonian (\ref{MKR}).\\
 
When  $U=1$,  the QE  spectrum  consists  just  of a  single,  infinitely
degenerate  level.   The natural  question  is  then precisely  how  this
infinite degeneracy is removed by slightly perturbing $\eta_j$ near their
integer values $m_j$.  For definiteness,  the perturbation of $\tau_j$ in
(\ref{tauj}) will be  made by perturbing $I$ near  $I=I_0$, leaving $t_j$
fixed.  Denoting  by $\epsilon$  the corresponding perturbation  in $\tau
=\hbar T/2I$, the perturbation in $\tau_j$ is given by
\begin{equation}\label{epj}
\epsilon_j = \frac{t_{j+1}-t_j}{T}\epsilon \ .
\end{equation}
Using the operator identity \cite{wil}
\[
e^A B e^{-A} = B + [A,\ B] + \frac{1}{2!}[A,\ [A,\ B]] +
                   \frac{1}{3!}[A,\ [A,\ [A,\ B]]] + \cdots \ ,
\]
and  formally expanding  the operators  $\exp (-i\tau_j{\hat  n}^2)= \exp
(-i\epsilon_j {\hat n}^2)$  in powers of $\epsilon_j$, we  find, to first
order in $\epsilon_j$,
\begin{equation}\label{U1}
U\approx 1 - \sum_{j=0}^{M-1}\epsilon_j
\left \{ i{\hat n}^2 - d_j k [2iV'(\theta ) {\hat n}+V''(\theta )] +
id_j^2 k^2 V^{\prime 2}(\theta )\right \} \ ,
\end{equation}
where   $d_j=\sum_{s=0}^j   c_s$,   and   the  prime   on   $V$   denotes
differentiation  with  respect  to   $\theta$.   Using  (\ref{epj}),  the
expression in (\ref{U1}) can be written, to first order in $\epsilon$, as
$\exp (-i\epsilon G_1)$, where
\begin{equation}\label{G1}
G_1 = \left [ {\hat n} -k{\bar d}V'(\theta )\right ]^2 +
k_{eff}^2 V^{\prime 2}(\theta )
\end{equation}
and  $k_{eff}=k\Delta   d$.   Here  ${\bar   d}$  and  $\Delta   d$  are,
respectively,  the   average  and   standard  deviation  of   $d_j$  with
``probability distribution" $(t_{j+1}-t_j)/T$:
\begin{equation}\label{avb}
{\bar d}=\sum_{j=0}^{M-1}\frac{t_{j+1}-t_j}{T} d_j \ , \ \ \ \ \
(\Delta d)^2=\sum_{j=0}^{M-1}\frac{t_{j+1}-t_j}{T} d_j^2 -{\bar d}^2\ .
\end{equation}
Using (\ref{avb})  and the definition of  $d_j$, it is easy  to show that
$k_{eff}$ (or $\Delta  d$) is invariant under cyclic  permutations of the
sequence $c_j$ \cite{note}.\\
 
Assuming for the moment the validity  of the expansion above in powers of
$\epsilon$ (see the discussion below), the  QE states $\psi$ in the limit
of infinitesimal $\epsilon$ are precisely the eigenstates of $G_1$,
\begin{equation}\label{eig}
G_1\psi = g\psi \ ,
\end{equation}
with  quasienergies  $\omega  =\epsilon g$  ($\epsilon  \rightarrow  0$).
Performing on (\ref{eig}) the gauge transformation
\begin{equation}\label{ga}
\varphi = \exp \left [ -ik{\bar d}V(\theta )\right ] \psi \ ,
\end{equation}
we obtain for $\varphi$, using (\ref{G1}), the eigenvalue equation
\begin{equation}\label{eq}
-\frac{d^2\varphi}{d\theta ^2} +
k_{eff}^2V^{\prime 2}(\theta )\varphi = g\varphi \ .
\end{equation}
We thus see that the QE problem for infinitesimal $\epsilon$ is just that
of a Schr\"{o}dinger equation (\ref{eq})  with a periodic potential.  The
spectrum  $g$ then  has a  band structure,  but because  of the  periodic
boundary condition  $\varphi (2\pi  )=\varphi (0)$,  only the  level with
zero quasimomentum is picked out from each band.  This gives, in general,
a point  spectrum.  Now,  being the solution  of the  linear differential
equation  (\ref{eq}), $\varphi  (\theta )$  is analytic  at least  in the
domain of analyticity  of $V'(\theta )$ \cite{inc}.  Let  $\gamma$ be the
smallest  distance  of a  singularity  of  $V'(\theta  )$ from  the  real
$\theta$-axis.  Then the Fourier-series  expansion of $\varphi (\theta )$
will  converge at  least within  an  infinite horizontal  strip of  width
$2\gamma$,  symmetrically   positioned  around  the   real  $\theta$-axis
\cite{inc}.  It  follows that the  Fourier coefficients of  $\varphi$ and
$\psi$  in (\ref{ga})  decay asymptotically  at least  as $\exp  (-\gamma
\vert n\vert)$.  This  means that in the immediate vicinity  of QAR there
takes  place {\it  asymptotic  exponential localization}  in the  angular
momentum  $n\hbar$,  with localization  length  $\xi_0$  not larger  than
$1/\gamma$.  In general, $\xi_0$ is determined entirely by the analytical
properties of $V'(\theta )$ (see examples in next section).\\
 
This  exponential   ``QAR-localization"  in   $L$,  following   from  Eq.
(\ref{eq}), justifies {\it a posteriori} the expansion above in powers of
$\epsilon$.  In fact,  the general expansion for $U$  in (\ref{UMKR}) can
be  formally  written  as  $\exp  (-iG)$,  $G=\sum_{j=1}^{\infty}\epsilon
^jG_j$.  Here the  Hermitian operators $G_j$ are polynomials  in $\hat n$
and  derivatives of  $V(\theta  )$ of  order not  larger  than $2j$  [the
leading  operator, $G_1$,  is given  by (\ref{G1})].   Thus, the  highest
power  of $\hat  n$ contributed  by $G_j$  appears in  this expansion  as
$(\epsilon  {\hat n}^2)^j$.   This  means that  by  choosing $\epsilon  <
n_{max}^{-2}$, where $n_{max}\gg 1/\gamma$, the eigenstates of $U$ (i.e.,
the QE states)  should be very close  to those of $G_1$,  at least within
the localization domain.   In the limit of  infinitesimal $\epsilon$, the
QE states should coincide with  the eigenstates of $G_1$.  The derivation
of Eq. (\ref{eq}) appears then to be self-consistent.\\
 
We now  show that  the effective Hamiltonian  (\ref{G1}) has  a classical
counterpart in the limit of very  small values of the quantity $T/I=2\tau
/\hbar$.  This limit  corresponds to the case of  infinitesimal values of
$\tau_j$ in (\ref{tauj}), i.e., infinitesimally  close to the special QAR
point $\eta  =\tau /2\pi =0$ ($m_j=0$  for all $j$).  Consider  the $M$th
iteration  of  the classical  map  (\ref{map}),  giving the  map  $(L_s,\
\theta_s)\rightarrow  (L_{s+M},\  \theta_{s+M})$.  Taking  carefully  the
limit $T\rightarrow 0$  in this map at fixed  $(t_{j+1}-t_j)/T$ and using
the condition (\ref{s0}), we obtain,  after a straightforward but tedious
calculation, the Hamilton equations
\begin{equation}\label{ehe}
\frac{dL}{dt} = -\frac{\partial H_{eff}}{\partial \theta}\ ,\ \ \ \ \
\frac{d\theta}{dt} = \frac{\partial H_{eff}}{\partial L}\ ,
\end{equation}
where
\begin{equation}\label{Hef}
H_{eff} = \frac{\hbar ^2}{2I} G_1 =
\frac{1}{2I}\left [ L -{\hat k}{\bar d}V'(\theta )\right ]^2 +
\frac{({\hat k}\Delta d)^2}{2I} V^{\prime 2}(\theta ) \ .
\end{equation}
Eqs. (\ref{ehe})  and (\ref{Hef}) show  that the general  MKR Hamiltonian
(\ref{MKR}), with coefficients $c_j$ satisfying (\ref{s0}), is integrable
in  the  limit $T\rightarrow  0$,  as  it  reduces  precisely to  the  1D
effective  Hamiltonian (\ref{Hef}).   The latter  is essentially  the QAR
effective   Hamiltonian    ({\ref{G1}),   and,   after    the   canonical
transformation $L'=L-{\hat k}{\bar d}V'(\theta )$ [analogous to the gauge
transformation  (\ref{ga})], it  becomes essentially  the Schr\"{o}dinger
Hamiltonian in  (\ref{eq}).  Thus, QAR-localization in  the infinitesimal
vicinity of $\eta =0$ is  associated with a classically integrable limit.
As $T$  is increased from  $0$, keeping the  quantities $(t_{j+1}-t_j)/T$
fixed  at some  rational  values  $m_j/m$ ($m=\sum_{j=0}^{M-1}m_j$),  the
QAR-localization  for   $\eta  =0$   will  repeat  periodically   in  the
infinitesimal vicinity  of $\eta =rm$,  for all integers $r$.   For these
values of  $\eta$, which are  equivalent to  $\eta =0$ but  correspond to
nonintegrable systems in a strong quantum regime, the QAR-localization is
only a  ``reflection" of  the classically integrable  limit $T\rightarrow
0$.  In  Appendix A,  we show that  QARs of periods  $p=1$ and  $p=2$ can
occur, in  general, if  $\tau_j$ in  (\ref{tauj}) is  an odd  multiple of
$\pi$.  Such values  of $\tau_j$ are {\it not}  equivalent to $\tau_j=0$,
since $\tau_j \bmod 2\pi \neq 0$.   In this case, the QAR-localization is
not even a reflection of a classically integrable limit.\\
 
It  is important  to notice  that the  limit $T\rightarrow  0$ (or  $\eta
\rightarrow 0$) at fixed ${\hat k}$ (or $k$) is {\it not} a semiclassical
limit.  In fact,  if the quantities $(t_{j+1}-t_j)/T$ are  kept fixed and
the  coordinate  transformation  $L'=(T/I)L$  is  performed  in  the  map
(\ref{map}), it becomes clear that the classical dynamics depends only on
the parameter $K=(T/I){\hat k}=2\tau k$.  The semiclassical limit is then
$\eta \rightarrow 0$ at fixed $K$,  not at fixed ${\hat k}$.  However, at
fixed $k\gg 1$, small  values of $\eta$ such that $K\ll  1$ may be viewed
as corresponding to a semiclassical regime of almost integrability.\\
 
 
\begin{center} {\bf IV. EXACTLY-SOLVABLE CASES}\\ \end{center}
 
In this  section we consider cases  of potentials $V(\theta )$  for which
the QE problem in the infinitesimal  vicinity of QAR [Eq. (\ref{eq})] can
be  solved in  closed form,  or at  least an  explicit expression  can be
obtained  for the  QAR-localization  length $\xi_0$.   Using these  exact
results, we shall provide strong  numerical evidence that the QE spectrum
and localization  features sufficiently close  to QAR are  well accounted
for by  the QAR effective Hamiltonian  (\ref{G1}).  As shown in  the next
section,  the  MKR dynamical  problem  can  be mapped  into  pseudorandom
tight-binding models, so that dynamical  localization (DL) is expected to
occur if $\eta =\tau /2\pi$ is sufficiently irrational.  If, in addition,
$\eta  -\eta_0$  is  small  enough,   this  DL  should  look  similar  to
QAR-localization.\\
 
Our first example is the  standard potential $V(\theta )=\cos (\theta )$,
for  which  Eq.   (\ref{eq})  reduces  to  the   {\it  Mathieu  equation}
\cite{inc,abra}
\begin{equation}\label{ma}
y''+\left [ a-2q\cos (2\theta )\right ] y=0 \ ,
\end{equation}
where $y=\varphi$, $a=g-(k\Delta d)^2/2$,  and $q=-(k\Delta d)^2/4$.  The
problem is then exactly solved in terms of the periodic Mathieu functions
$y=ce_r(\theta   ,\   q)$   (symmetric)   and   $y=se_r(\theta   ,\   q)$
(anti-symmetric),   with   corresponding   eigenvalues   $a=a_r(q)$   and
$a=b_r(q)$.  Explicit expressions for these functions and eigenvalues, as
well as a detailed discussion of  their properties, can be found in Refs.
\cite{inc,abra}.  From  Eq. (\ref{ga}) the Fourier  coefficients $\psi_n$
and  $y_n$  of  $\psi$  and $y=\varphi$,  respectively,  are  related  by
$\psi_n=\sum_{j}i^jJ_j(k{\bar  d})y_{n-j}$, where  $J_j(k{\bar d})$  is a
Bessel function.  Since  the dominant decay rate of both  $J_n$ and $y_n$
with $n$  is like $n^{-n}$ \cite{abra},  this is also the  dominant decay
rate of  $\psi_n$.  This  strong localization in  $L$ space,  faster than
exponential,  could  be expected  from  the  fact that  $V(\theta  )=\cos
(\theta  )$  is  an  entire function  (analyticity-strip  width  $2\gamma
=\infty$), so  that the  asymptotic localization length  $\xi_0 =1/\gamma
=0$.\\
 
Let us now check to what  extent the QAR effective Hamiltonian (\ref{G1})
for  $V(\theta  )=\cos  (\theta  )$  reproduces  accurately  the  quantum
dynamics and  QE spectrum for $\eta  -\eta_0=\epsilon /2\pi$ sufficiently
small and  irrational.  We  have studied numerically  the case  of $M=3$,
with $c_0=c_1=1$,  $c_2=-2$, and  $\tau_j=\tau /3$  for all  $j$ (chaotic
orbits  for this  system are  shown  in Fig.\  \ref{logl}).  The  quantum
dynamics  of a  wave-packet  initially equal  to  $\vert n=0\rangle$  was
investigated using  a basis of  up to 512 angular-momentum  states around
$n=0$.   The   wave-packet  was  propagated  in   time  using  well-known
algorithms \cite{fish}.  In Fig.\  \ref{coll}, we plot the kinetic-energy
expectation  value, $E_0=\langle  L^2/2I\rangle$,  as a  function of  the
``real time" $t=\epsilon  s$ ($s$ is the number of  applications of $U$),
for $k=2.0$ ($k_{eff}\equiv k\Delta d=\sqrt{8/3}$) and several irrational
values of  $\epsilon /2\pi$.  We observe  that almost all the  data falls
close  to the  same curve,  even  for values  of $\epsilon$  as large  as
$\epsilon =0.28$.  For  smaller values of $k_{eff}$, the  data falls much
more  accurately on  the same  curve (see  an example  for $M=2$  in Ref.
\cite{des}).  This is evidence that,  even for $\epsilon$ not very small,
the  quantum  dynamics over  a  significant  time  interval can  be  well
described  by the  approximate evolution  operator $\overline{U}^s=  \exp
(-i\epsilon  sG_1)=\exp   (-itG_1)$,  generated  by  the   QAR  effective
Hamiltonian $G_1$.   In Fig.\ \ref{freq},  we plot the  Fourier transform
$E_0(\nu )$  of $E_0(t)$ for $\epsilon  =0.1$ and several values  of $k$.
The positions  of the  various peaks  in $E_0(\nu  )$ must  correspond to
values of  $\epsilon \nu$  equal to  the spacings  between QE  levels.  A
comparison with  the level spacings  corresponding to eigenvalues  of the
Mathieu equation  (\ref{ma}) shows  excellent agreement.  This  is strong
evidence that  sufficiently close  to QAR  the QE  spectrum is  very well
accounted for by the QAR operator $G_1$ [i.e., Eq. (\ref{eq})].\\
 
Dynamical localization (DL) \cite{not} in the vicinity of QAR will now be
compared  with   QAR-localization  (we   assume  in  what   follows  that
$\eta_0=0$).  For  this purpose,  it will be  sufficient to  consider the
localization  of steady-state  probability  distributions $\langle  \vert
\psi  (n)  \vert  ^2\rangle$  over angular  momentum  $n\hbar$.   Such  a
distribution  is calculated  by propagating  an initial  angular-momentum
state  $\vert n=0\rangle$  for  a sufficiently  long  time interval,  and
averaging  then  the  results  for  $\vert \psi  (n)  \vert  ^2$  over  a
subinterval at the end of  this interval.  All our numerical calculations
in what follows have been restricted to  the simple case of the $M=2$ MKR
with  $c_0=-c_1=1$ and  $t_1=T/2$  (the  ``two-sided" KR  \cite{us,des}).
Fig. 4  shows a semi-logarithmic  plot of  $\langle \vert \psi  (n) \vert
^2\rangle$ for $V(\theta )=\cos (\theta  )$, $k=10$, and three irrational
values of $\eta  =\tau /2\pi$, $\tau =10^{-5},\  10^{-3},\ 10^{-1}$.  For
the first two  values of $\tau$, corresponding to close  vicinity of QAR,
$\langle \vert  \psi (n)  \vert ^2\rangle$ appears  to decay  faster than
exponential,  in  accord  with  the solutions  of  the  Mathieu  equation
(\ref{ma}) (i.e., QAR-localization).  On the other hand, the localization
for  $\tau  =10^{-1}$ looks  quite  different  from QAR-localization;  it
appears,  in  fact, to  be  almost  exponential.   These results  can  be
explained as follows.  The  classical nonintegrability parameter is given
by $K=(T/I){\hat k}=2\tau k$ (see previous section).  Thus, the first two
values of $\tau$  correspond to a semiclassical regime  (small $\tau$) of
almost integrability (small $K$), and  the observed DL \cite{not} is then
similar to  QAR-localization, as predicted from  the integrable effective
Hamiltonian  (\ref{G1}) (however,  a basic  difference between  these two
localizations  is expected,  as  pointed  out in  Sec.   VI).  For  $\tau
=10^{-1}$, on  the other  hand, $K=2$,  corresponding to  a semiclassical
regime  (small $\tau$)  of  global  chaotic diffusion  in  the $M=2$  MKR
\cite{us}.  As in  the KR case \cite{she,blu},  the localization observed
here should be  characterized by a DL length  $\xi$ completely determined
by the  chaotic-diffusion coefficient  $D$, $\xi  \approx D/2$  (see Sec.
VI).  This is  in contrast with QAR-localization, which  is determined by
the  analytical  properties  of   the  potential.   The  transition  from
QAR-localization to DL, as $\tau$ is  varied in a semiclassical regime at
fixed $k$, will be studied in more detail in Sec.  VI.\\
 
Our second example is the potential
\begin{equation}\label{atan}
V(\theta ) = A \arctan [\kappa \cos (\theta)-\kappa_0] \ ,
\end{equation}
where   $A$,  $\kappa$,   and   $\kappa_0$  are   some  constants.    The
QAR-localization  length  $\xi_0$  for  (\ref{atan})  can  be  determined
exactly  as follows.   The function  $V'(\theta )$  assumes simple  poles
$\theta_0$  satisfying the  equation $\kappa\cos  (\theta_0)-\kappa_0=\pm
i$.  The  distance of any of  these poles from the  real $\theta$-axis is
$\gamma =\vert\mbox{Im}(\theta_0)\vert$, and we easily find that
\begin{equation}\label{gamma}
2\kappa \cosh (\gamma ) =
[1+(\kappa_0+\kappa )^2]^{1/2} + [1+(\kappa_0-\kappa )^2]^{1/2}\ .
\end{equation}
Consider  now the  Fourier-series  expansion for  the solutions  $\varphi
(\theta )$ of Eq. (\ref{eq}).  We  claim that in the case of (\ref{atan})
the Fourier  coefficients $\varphi_n$ must decay  {\it asymptotically} as
$\varphi_n \asymp  \exp (-\gamma  \vert n\vert )$.   The QAR-localization
length  is then  $\xi_0=1/\gamma$.  To  show  this, we  observe that  the
simple  poles  $\theta_0$  correspond   to  {\it  regular  singularities}
\cite{inc} of  order $2$ of  Eq. (\ref{eq}).  The exponents  $\rho_1$ and
$\rho_2$ for these singularities are easily determined from the quadratic
``indicial"  equation  \cite{inc}  for  (\ref{eq}):  $\rho_1=\rho_2=1/2$.
Since the exponents are equal,  the general solution of (\ref{eq}) around
$\theta =\theta_0$ assumes the form \cite{inc}
\begin{equation}\label{sol}
\varphi (\theta  ) = (\theta -\theta_0)^{1/2} \{ R_1(\theta -\theta_0) +
R_2(\theta -\theta_0) [b_1+b_2\log (\theta -\theta_0)]\} \ ,
\end{equation}
where $R_1(\theta )$ and $R_2(\theta )$ are analytic (can be expressed as
Taylor expansions)  around $\theta  =\theta_0$, and  $b_1$ and  $b_2$ are
arbitrary  constants.    It  follows   from  (\ref{sol})  that   all  the
derivatives of $\varphi (\theta )$ diverge at $\theta =\theta_0$.  Let us
now continue  the Fourier-series expansion  for $\varphi (\theta  )$ into
the   complex   $\theta$   plane.     Defining   the   complex   variable
$z=e^{i\theta}$, one gets  a Laurent expansion in $z$  which converges at
least  in  a  ``ring"  excluding  the  singularities  of  Eq.  (\ref{eq})
\cite{inc},    i.e.,   for    $\vert   \mbox{Im}(\theta    )   \vert    <
\mbox{Im}(\theta_0)=\gamma$.   However,  since  all  the  derivatives  of
(\ref{sol}) diverge as  $\theta \rightarrow \theta_0$, this  must be also
the  case for  the derivatives  of the  Laurent expansion.   By a  simple
consideration of the latter  derivatives, the desired relation $\varphi_n
\asymp \exp (-\gamma \vert n\vert )$ is obtained.\\
 
Figs. 5  and 6  show semi-logarithmic  plots of  $\langle \vert  \psi (n)
\vert ^2\rangle$ for the $M=2$ MKR  (defined as above) with the potential
(\ref{atan}) ($A=1$, $\kappa=1$, and $\kappa_0=0$) and for several values
of $k$ and $\tau$.  The straight  line (dotted) in both figures has slope
$-2\gamma$,  where $\gamma$  is determined  from Eq.  (\ref{gamma}).  For
$\tau$ sufficiently  small, this  slope is  expected to  be close  to the
asymptotic rate  of exponential  decay of $\langle  \vert \psi  (n) \vert
^2\rangle$.  We see  that this is indeed the case  whenever the classical
parameter $K=2\tau  k$ is  small enough,  $K<1$, corresponding  to almost
integrability  or local  chaos.  In  particular, for  $k=2$ (Fig.  5) the
decay  rate appears  to be  equal to  $2\gamma$ for  all three  values of
$\tau$.  In fact, for $k=2$ and $A=1$ the potential (\ref{atan}) leads to
a nearest-neighbor pseudorandom tight-binding model whose DL length $\xi$
appears to be  independent of $\tau$ (see  Sec.  VI and Fig.  8).  On the
other hand, for $k=10$  (Fig. 6) the DL length is  quite sensitive to the
value of  $\tau$.  In  the case  of $\tau  =10^{-1}$, corresponding  to a
semiclassical regime of global chaotic diffusion ($K=2$), the observed DL
looks quite different from QAR-localization, with an asymptotic DL length
$\xi \approx  D/2 \neq  \xi_0$.  This  case is similar  to that  of $\tau
=10^{-1}$ in Fig. 4.\\
 
\begin{center}  {\bf   V.   MULTI-CHANNEL   PSEUDORANDOM\\  TIGHT-BINDING
MODELS}\\ \end{center}
 
We now show how the MKR  dynamical problem with (\ref{MKR}) can be mapped
into a  pseudorandom tight-binding model,  in analogy to the  ordinary KR
case \cite{fish}.   For simplicity, we  shall assume that  the quantities
(\ref{tauj}) are  all equal,  $\tau_j =\tau  /M$ for  all $j$  (i.e., the
kicks  are equidistant  in time).   Let $u_j^{\pm}(\theta  )$, $j=0,...,\
M-1$, denote a  QE state with quasienergy $\omega$ at  time $t=t_j\pm 0$.
The following relations hold:
\begin{equation}\label{dy1}
u_j^{+}(\theta ) = \exp [-ic_jkV(\theta )] u_j^{-}\ ,
\end{equation}
\begin{equation}\label{dy2}
u_{j,n}^{-}=e^{-i\tau n^2/M}u_{j-1,n}^{+}\ (0 < j \leq M-1)\ ,\ \ \ \
u_{0,n}^{-}=e^{i(\omega -\tau n^2/M)}u_{M-1,n}^{+}\ ,
\end{equation}
where $u_{j,n}^{\pm}$ is the  $L$-representation of $u_j^{\pm}(\theta )$.
We define, in some analogy with Ref. \cite{fish},
\begin{equation}\label{uW}
u_j(\theta )=
e^{ij\omega /M}\frac{u_j^{+}(\theta )+u_j^{-}(\theta )}{2}\ ,\ \ \ \ \
e^{-ic_jkV(\theta )}=\frac{1+iW_j(\theta )}{1-iW_j(\theta )}\ ,
\end{equation}
so that $W_j(\theta )=-\tan [c_jkV(\theta )/2]$.  Simple manipulations of
Rels.  (\ref{dy1}), (\ref{dy2}),  and (\ref{uW})  yield a  system of  $M$
equations  for  the  $L$-representation   $u_{j,n}$  of  $u_j(\theta  )$,
$j=0,...,\ M-1$:
\begin{equation}\label{sys}
u_{j+1,n}-i\sum_r W_{j+1,n-r}u_{j+1,r} = e^{i(\omega -\tau n^2)/M}
\left ( u_{j,n}+i\sum_r W_{j,n-r}u_{j,r}\right ) \ ,
\end{equation}
where $u_{M,n}\equiv u_{0,n}$, and $W_{j,n}$ is the $L$-representation of
$W_j(\theta )$.  Unless otherwise specified, the index $r$ in (\ref{sys})
runs  over all  the integers.   We now  introduce the  Fourier transforms
${\bar  u}_{s,n}$ and  $\overline{W}_{s,n}$ of  $u_{j,n}$ and  $W_{j,n}$,
respectively, in the variable $j$:
\begin{equation}\label{fouu}
{\bar u}_{s,n} = \frac{1}{M} \sum_{j=0}^{M-1} u_{j,n} e^{2\pi ijs/M}\ ,
\end{equation}
\begin{equation}\label{fouW}
\overline{W}_{s,n}=\frac{1}{M}\sum_{j=0}^{M-1}W_{j,n} e^{2\pi ijs/M}\ .
\end{equation}
Using the  expressions (\ref{fouu})  and (\ref{fouW}) in  (\ref{sys}), we
obtain, after simple algebraic manipulations,
\begin{equation}\label{tbm}
T_n^{(s)}{\bar u}_{s,n}+\sum_{r\neq n}\overline{W}_{0,n-r}{\bar u}_{s,r}
+ \sum_r\sum_{s'\neq s}\overline{W}_{s-s',n-r}{\bar u}_{s',r}
= E {\bar u}_{s,n}\ ,
\end{equation}
where, for $s=0,...,\ M-1$,
\begin{equation}\label{Tns}
T_n^{(s)} = -\tan [(\tau n^2 - 2\pi s - \omega )/2M]\ ,
\end{equation}
the index  $s'$ takes all  the integer  values $s'=0,...,\ M-1$  with the
exception  of  $s'=s$,  and  $E=-\overline{W}_{0,0}$.   Eqs.  (\ref{tbm})
describe a tight-binding  model of an $M$-channel  strip \cite{der}.  The
on-site  potential in  channel $s$  is  given by  $T_n^{(s)}$, while  the
hopping constants within a channel are $\overline{W}_{0,n}$.  The hopping
constants  from  channel   $s$  to  channel  $s'\neq  s$   are  given  by
$\overline{W}_{s-s',n}$.\\
 
A particularly interesting case arises when $c_j$ assumes only the values
$0,\ \pm c$ for some constant $c$,  which will be chosen, without loss of
generality,  equal  to 1.   From  Rel.  (\ref{fouW}) and  the  definition
$W_j(\theta )=-\tan [c_jkV(\theta )/2]$, it follows then that
\begin{equation}\label{sinW}
\overline{W}_{s,n} = c(s)W_n \equiv
\frac{1}{M} \sum_{j=0}^{M-1} c_j e^{2\pi ijs/M} W_n\ ,
\end{equation}
where   $c(s)$   is   defined   by  (\ref{sinW})   and   $W_n$   is   the
$L$-representation of  $W(\theta )\equiv  -\tan [kV(\theta  )/2]$.  Rels.
(\ref{sinW})  and  (\ref{s0})  imply   now  that  the  hopping  constants
$\overline{W}_{0,n}$ within a channel are identically zero, including, of
course, $E=-\overline{W}_{0,0}$.  In this  {\it singular} case, the model
(\ref{tbm})  looses much  of its  physical  meaning.  In  Appendix C,  we
consider in some  detail nearly-singular cases of  (\ref{tbm}), for which
the hopping constants $\overline{W}_{0,n}$ are nonzero but small.  Notice
that the $M=2$ case is always singular, since $c_0=-c_1$ from (\ref{s0}).
In  this  case,  however,   a  physically  meaningful  two-channel  model
\cite{des} can  be derived  directly from  Eqs. (\ref{sys}),  without the
need  of the  Fourier transforms  (\ref{fouu}) and  (\ref{fouW}).  Simple
manipulations of Eqs. (\ref{sys}) give in this case:
\begin{equation}\label{And}
\begin{array}{rcl}
T_nu_{0,n}+S_nu_{1,n}+\sum _{r\neq 0}W_{n-r}u_{0,r} & = & Eu_{0,n}\ ,
\\
-T_nu_{1,n}-S_nu_{0,n}+\sum _{r\neq 0}W_{n-r}u_{1,r} & = & Eu_{1,n}\ ,
\end{array}
\end{equation}
where  $T_n=\cot  (\zeta_n)$,   $S_n=-1/\sin  (\zeta_n)$,  $\zeta_n=(\tau
n^2-\omega  )/2$,  and  $E=-W_0$.   The  on-site  potential  and  hopping
constants within each  channel are, respectively, $T_n$  and $W_n$, while
$S_n$ are the coupling constants between  the channels.  As shown in Sec.
VI, the  $M=2$ MKR  is essentially  equivalent to an  ordinary KR  if the
potential satisfies  $V(\theta +\pi  )=-V(\theta )$.   In this  case, the
dynamical problem  can be  conveniently approached  using the  well known
tight-binding models for the KR \cite{fish,she}.\\
 
For irrational $\eta  =\tau /2\pi$, $T_n^{(s)}$ in  (\ref{Tns}) [or $T_n$
and  $S_n$ in  (\ref{And})]  is a  pseudorandom sequence  \cite{fish,gri}
which, by arguments similar to those used in the KR case \cite{fish}, may
lead to Anderson-like localization of  the eigenstates of (\ref{tbm}) [or
of  (\ref{And})].    This  is  the  {\it   dynamical  localization}  (DL)
\cite{not} of the  corresponding QE states in  angular momentum.  Another
source of  randomness, which does  not involve $\eta$ but  may contribute
significantly  to  DL,  is  the   dependence  of  the  hopping  constants
$\overline{W}_{s,n}$ on the distance  $s$ between channels, especially in
the limit of  very large $M$.  This may reflect  a possible randomness of
the sequence $c_j$ in the ``time index" $j$.\\
 
Clearly,  the  pseudorandomness  is  not  defined  in  the  infinitesimal
vicinity  of QAR  (infinitesimal $\epsilon  =\tau \bmod  2\pi$), so  that
QAR-localization is not DL, strictly speaking.  However, when approaching
a  QAR point  $\eta_0$, the  pseudorandomness is  guaranteed by  choosing
$\epsilon/2\pi$ in a sequence of ``strong" irrationals $\epsilon_l/2\pi$,
$l=1,\  2,...$, converging  to 0  [e.g., $\epsilon_l=2\pi/(l+\varrho  )$,
where $\varrho$ is the golden mean].   It is then quite possible that for
$\epsilon =\epsilon_l$  the model (\ref{tbm}) [or  (\ref{And})] will have
exponentially-localized  eigenstates.  In  fact,  the numerical  evidence
presented  in  the previous  section  (see  also next  section)  strongly
indicates that this is indeed the case for $l$ sufficiently large.  Then,
provided  the  QAR condition  (\ref{s0})  is  satisfied, the  quasienergy
$\omega$  in  (\ref{Tns})  should   be  given  approximately  by  $\omega
=\epsilon g$, where $g$ is an eigenvalue of the leading operator $G_1$ in
(\ref{G1}).  Moreover, in  the $M=2$ case of (\ref{And}),  one can easily
determine, using  (\ref{dy1}), (\ref{uW}),  and (\ref{ga}),  the accurate
relation  expected   between  the   solutions  of  Eqs.   (\ref{eq})  and
(\ref{And}) for  small $\epsilon$:  $u_0(\theta )\approx  \cos [kV(\theta
)/2]\varphi (\theta )$.\\
 
For rational  values of  $\eta =m/p$  ($m$ and  $p$ are  relatively prime
integers), the on-site potential $T_n^{(s)}$  in (\ref{Tns}) [or $T_n$ in
(\ref{And})]   is  periodic   in   $n$  with   period   $p$.   Then,   by
straightforward application  of Bloch theorem,  one can easily  show that
the  QE   spectrum  $\omega$  is,  in   general,  absolutely  continuous,
consisting of  $p$ bands.   This corresponds  to usual  quantum resonance
\cite{izr}.  However, when  the QAR conditions are  satisfied, either for
$p=1$ or $p=2$ [see (\ref{s0}) and Appendix A], a QE band ``collapses" to
an infinitely degenerate level.\\
 
\begin{center} {\bf VI.  TRANSITION  FROM QAR-LOCALIZATION TO\\ DYNAMICAL
LOCALIZATION IN A SEMICLASSICAL REGIME}\\ \end{center}
 
In this section,  we study the dependence  of the DL length  $\xi$ on the
distance $\eta -\eta_0$  from the QAR point $\eta_0=0$.   This study will
be  performed in  a  natural way  at  fixed $k$,  so  that $\eta_0=0$  is
associated with a  classically integrable limit (see Sec.   III).  If $k$
is sufficiently  large, the classical  parameter $K=2\tau k=4\pi  \eta k$
may correspond  to a case of  global chaotic diffusion already  for small
values of $\eta <1$, i.e.,  in the semiclassical regime.  The approximate
relation $\xi  \approx D/2$  \cite{she,blu}, where  $D$ is  the diffusion
coefficient  [see definition  (\ref{D})  below], should  then hold.   The
transition from  QAR-localization to DL  in the semiclassical  regime, as
$\eta$ is  increased from $0$, can  be understood in a  most illuminating
way  if  the problem  is  approached  by  using a  suitable  pseudorandom
tight-binding model,  equivalent to the dynamical  system.  This approach
is also most convenient for an accurate calculation of $\xi$.\\
 
For  simplicity,   we  shall   restrict  ourselves   to  the   $M=2$  MKR
($c_0=-c_1=1$,  $t_1=T/2$)  with  potentials  satisfying  $V(\theta  +\pi
)=-V(\theta )$ (as in the numerical examples in Figs. 4-6).  We denote by
$U_{2KR}(\tau )$ the evolution operator for  this system at a given value
of  $\tau$.   Similarly,  we  denote by  $U_{KR}(\tau  )$  the  evolution
operator (\ref{UKR}) for  the ordinary KR.  Using  Rels. (\ref{UMKR}) and
(\ref{pi}) with $V(\theta  +\pi )=-V(\theta )$, it is easy  to derive the
following exact relation
\begin{equation}\label{U2KR}
U_{KR}^2(\pi +\epsilon /2)=U_{2KR}(\epsilon )\ .
\end{equation}
Rel. (\ref{U2KR})  means that the  quantum dynamics  of the $M=2$  MKR at
distance  $\epsilon$  from  the  QAR   point  $\tau  =0$  is  essentially
equivalent to  that of  the ordinary  KR at  distance $\epsilon  /2$ from
$\tau =\pi$.  The latter value of  $\tau$ is precisely a {\it period-two}
QAR point  for the KR  (see Sec.  II  and Appendix A).   This equivalence
between the  two systems  enables one  to study the  $M=2$ MKR  using the
simple tight-binding  models for the  KR \cite{fish,she}, instead  of the
two-channel model (\ref{And}).\\
 
Our basic potential is the standard  one, $V(\theta )=\cos (\theta )$.  A
convenient  tight-binding  model  for  the KR  with  this  potential  was
proposed by Shepelyansky \cite{she}:
\begin{equation}\label{tbKR}
\sum_{r=-\infty}^{\infty} J_r(k/2)\sin [(\tau n^2-\omega +\pi r)/2]
{\bar u}_{n+r} = 0\ ,
\end{equation}
where $J_r(k/2)$ is the Bessel function, $\omega$ is the quasienergy, and
${\bar u}_n$ is related to  the angular-momentum representation of the QE
states.   Using   Rel.  (\ref{U2KR})  in  (\ref{tbKR}),   we  obtain  the
corresponding tight-binding model for the $M=2$ MKR:
\begin{equation}\label{tb2KR}
\sum_{r=-\infty}^{\infty} J_r(k/2)
\sin \{ [(2\pi +\tau )n^2-\omega +2\pi r]/4\} {\bar u}_{n+r} = 0\ ,
\end{equation}
where all the quantities $\tau$, $\omega$,  and ${\bar u}_n$ now refer to
the $M=2$ MKR.  Let us briefly  recall how the asymptotic DL length $\xi$
can  be calculated  from such  tight-binding models  \cite{she,blu}.  The
hopping  constants  in  (\ref{tb2KR}),   given  by  the  Bessel  function
$J_r(k/2)$, decay faster than exponentially for $\vert r\vert >k/2$:
\begin{equation}\label{bes}
J_r(k/2) \sim \ \left \{
\begin{array}{ll}
(4/\pi k)^{1/2}\cos [k/2 -\pi r/2 -\pi /4)
& \mbox{\ \ \ \ \ for $\vert r\vert < k/2$,} \\
(1/2\pi \vert r\vert )^{1/2}(e k/4 r)^{\vert r\vert}
& \mbox{\ \ \ \ \ for $\vert r\vert > k/2$.}
\end{array}
\right .
\end{equation}
It is  therefore reasonable  to approximate (\ref{tb2KR})  by restricting
$r$ to  the finite range  $\vert r\vert  \leq N$, for  sufficiently large
$N$.  This truncated form of Eq. (\ref{tb2KR}) can be easily written as a
transfer-matrix problem \cite{she,blu}
\begin{equation}\label{tm}
\sigma_{s+1} = \Gamma_s \sigma_s \ ,
\end{equation}
where  $\sigma_s$   is  the   $2N$-dimensional  vector   with  components
$\sigma_s^{(r)}={\bar  u}_{s-r}$,  $r=-N+1,...,N$,  and $\Gamma_s$  is  a
$2N\times 2N$ symplectic  matrix.  One may interpret (\ref{tm})  as a map
describing a  Hamiltonian dynamical  system with  $N$ degrees  of freedom
\cite{lic}.   The  vector  $\sigma_0$  is  mapped  into  $\sigma_n$,  for
arbitrary $n > 0$, by the product matrix
\begin{equation}\label{sm}
\Lambda_n = \Gamma_{n-1} \Gamma_{n-2} \cdots \Gamma_0 \ .
\end{equation}
Since the  matrix (\ref{sm})  is, obviously, symplectic,  its eigenvalues
$\lambda   (n)$   come   in   $N$   reciprocal   pairs   $[\lambda_r(n),\
\lambda_r^{-1}(n)]$, $r=1,...,N$,  and we can always  assume the ordering
$1\leq \vert  \lambda_1(n)\vert \leq \vert \lambda_2(n)\vert  \leq \ldots
\vert \lambda_N(n)\vert$.   The {\it  minimal Lyapunov exponent}  for the
map (\ref{tm}),
\begin{equation}\label{mle}
\gamma_N =
\lim_{n\rightarrow \infty} \frac{1}{n} \ln \vert \lambda_1(n)\vert \ ,
\end{equation}
determines then  an $N$th-order  approximation $\xi_N=1/\gamma_N$  to the
asymptotic DL length $\xi$.  Since  an accurate calculation of $\gamma_N$
becomes extremely time-consuming as $N$ is increased, important questions
are  how fast  $\xi_N$  converges to  its  limit value  $\xi$  as $N$  is
increased,  and  whether  $\xi_N$ has  a  well-defined  quantum-dynamical
meaning {\it per se}.\\
 
We have therefore studied $\gamma_N$ as a function of both $N$ and $\eta$
near QAR.  The  value of $k$ was  fixed at $k=10$, and  several values of
$\tau$ were  considered in  the interval $10^{-6}\leq  \tau \leq  1$.  We
have calculated  $\gamma_N$ for $N\leq  15$, using the well  known method
\cite{she,blu,lic} for  determining the  Lyapunov spectra of  products of
matrices such as  (\ref{sm}).  The method is based  on direct application
of the map (\ref{tm}) a large  number $n=n_{max}$ of times, such that for
$n>n_{max}$ the  matrices $\Gamma_n$ can  be considered as  random.  This
randomness should  be realized  to some extent  by the  pseudorandom term
$\tau  n^2/4$ in  (\ref{tb2KR})  if $\tau  [(n+1)^2-n^2]/4\sim 2\pi$,  or
$n=n_{max}\sim  4\pi  /\tau$.   We  have used  $n_{max}=10^7$  for  $\tau
=10^{-6},\ 10^{-5}$, and $n_{max}=10^6$ for  the larger values of $\tau$.
We  have checked  that  this choice  of  $n_{max}$ yields  well-converged
results by calculating  and comparing $\gamma_N$ for  different values of
$n   \leq   n_{max}$,   e.g.,   $n=10^5,\   2\cdot   10^5,...,10^6$   for
$n_{max}=10^6$.   The  final  results  are   shown  in  Fig.  7  using  a
semi-logarithmic plot.   We observe that $\gamma_N$  clearly decreases in
the interval  $N\leq 3$ for all  values of $\tau$.  For  larger values of
$N$ and  for very  small $\tau$ (immediate  vicinity of  QAR), $\gamma_N$
appears  to increase  without  bounds  with $N$.   For  $\tau \geq  0.1$,
$\gamma_N$ appears to ``saturate" around some limit value which decreases
as  $\tau$  is increased.   For  $\tau  \sim  1$, the  saturation  occurs
immediately after $N>3$.\\
 
The apparently unbounded  increase of $\gamma_N$ with $N$  for very small
$\tau$ is consistent  with the fact that  the asymptotic QAR-localization
length for  $V(\theta )=\cos  (\theta )$ is  $\xi_0=0$ (analyticity-strip
width $2\gamma  =2\gamma_{\infty}=\infty$).  We now show  that $\gamma_N$
has a well-defined  quantum-dynamical meaning also for  finite $N$.  This
is because the truncated version of the tight-binding model (\ref{tb2KR})
turns out to be exactly equivalent  to the dynamical problem (KR or $M=2$
MKR) with a  potential $V_N(\theta )$ replacing  $V(\theta )=\cos (\theta
)$.   Then  $\xi_N=1/\gamma_N$  is  the DL  length  for  this  potential,
reducing to  the QAR-localization length  $\xi_{N,0}$ in the  limit $\tau
\rightarrow 0$.  For  very small $\tau$ (e.g., $\tau  =10^{-6}$ and $\tau
=10^{-5}$ in  Fig. 7),  $\xi_N$ should be  an excellent  approximation to
$\xi_{N,0}$.  The potential $V_N(\theta )$  can be easily determined from
the $g$-function approach of Shepelyansky \cite{she}:
\begin{equation}\label{WN}
W_N(\theta )=g_N(\theta )\exp [-ikV_N(\theta )/2]=
\sum_{r=-N}^N J_r(k/2) e^{ir(\theta -\pi /2)} \ ,
\end{equation}
where $g_N(\theta )$  is some real function.  Solving  Eq. (\ref{WN}) for
$V_N(\theta )$, we obtain
\begin{equation}\label{VN}
V_N(\theta )=-\frac{2}{k}\arctan \left [
\frac{\sum_{r=-N}^N J_r(k/2)\sin (r\theta -r\pi /2)}
{\sum_{r=-N}^N J_r(k/2)\cos (r\theta -r\pi /2)}\right ] \ .
\end{equation}
Using  $J_{-r}(k/2)=(-1)^rJ_r(k/2)$, it  is easy  to see  that (\ref{VN})
satisfies  $V_N(\theta +\pi)=-V_N(\theta  )$, so  that Rel.  (\ref{U2KR})
between the KR and the $M=2$  MKR holds for the potential $V_N(\theta )$.
For  $N=1$, (\ref{VN})  reduces  to (\ref{atan})  with $\kappa_0=0$  (see
below).  The QAR-localization length for  (\ref{VN}) can be determined as
in  the case  of the  potential (\ref{atan}).   It is  easily shown  that
$V_N'(\theta )$ assumes simple poles $\theta_0$ satisfying the equation
\[
W_N(\theta_0)=\sum_{r=-N}^N J_r(k/2) e^{ir(\theta_0 -\pi /2)}=0 \ ,
\]
or $g_N(\theta_0)=0$,  since the  exponential function in  (\ref{WN}) can
never vanish.  The QAR-localization length for $V_N(\theta )$ is given by
$\xi_{N,0}=1/\gamma_N^{(0)}$,  where  $\gamma_N^{(0)}$  is  the  smallest
distance of  a pole $\theta_0$ from  the real $\theta$-axis.  It  is also
the  half-width   of  the   strip  of   analyticity  of   $V_N(\theta  )$
(Fourier-series  representation).  Since  $\gamma =\infty$  for $V(\theta
)=\cos (\theta )$,  $\lim_{N\rightarrow \infty}\gamma_N^{(0)}=\infty$, so
that $\gamma_N^{(0)}$ must generally increase  with $N$, as is clear from
Fig. 7 for very small $\tau$.\\
 
We thus see that in the limit $\tau \rightarrow 0$ $\gamma_N$ is both the
minimal Lyapunov  exponent of a  $2N\times 2N$ symplectic matrix  and the
half-width  of  analyticity  of  the potential  $V_N(\theta  )$  for  the
truncated  problem.   As $N$  increases,  the  region of  analyticity  of
$V_N(\theta )$ increases without  bounds, and the QAR-localization length
$\xi_{N,0}\rightarrow 0$.  With  this in mind, we now  consider the cases
where $\gamma_N$  appears to saturate  for $N$ not  too large in  Fig. 7.
These  cases correspond  to regimes  of local  ($K<1$) or  global ($K>1$)
chaos.   As  in  the  KR  case \cite{she,blu},  we  expect  that  in  the
global-chaos   regime   the   saturated    value   of   the   DL   length
$\xi_N=1/\gamma_N$ is given approximately by $\xi \approx D/2$, where $D$
is the classical chaotic diffusion coefficient.  In complete analogy with
Refs. \cite{rot,she}, we shall define $D$  for the MKR map (\ref{map}) as
follows:
\begin{equation}\label{D}
D = \lim_{r\rightarrow \infty}\frac{\langle (L_{rM}-L_0)^2\rangle}{rM}\ ,
\end{equation}
where $\langle  \ \rangle$  denotes average over  an ensemble  of initial
conditions $\{  (L_0,\ \theta_0 )\}$.   In a strong-chaos regime,  we may
assume, as usual, that the  angles $\theta_s$ are independent and random,
giving       vanishing      force-force       correlations,      $\langle
V'(\theta_s)V'(\theta_{s'})\rangle  =0$.    Using  then   (\ref{map})  in
(\ref{D}), we obtain the following expression for $D$:
\begin{equation}\label{D0}
D=\frac{k^2}{M}\lim_{r\rightarrow \infty}\frac{1}{r}\left \langle
\left [ \sum_{r'=0}^{r-1}\sum_{j=0}^{M-1}c_jV'(\theta_{r'M+j})\right ]^2
\right \rangle = k^2\langle c^2\rangle
\int_{0}^{2\pi} V^{\prime 2}(\theta )\frac{d\theta}{2\pi}\ ,
\end{equation}
where  $\langle c^2\rangle  =\sum_{j=0}^{M-1}c_j^2/M$,  and,  as in  Ref.
\cite{she}, we use units such that  $\hbar =1$, giving ${\hat k}=k$.  For
the $M=2$  MKR $\langle  c^2\rangle =1$, and,  with $k=10$  and $V(\theta
)=\cos  (\theta )$,  we find  from (\ref{D0})  that $D=50$.   The minimal
saturated value  of $\gamma_N$ in  Fig. 7 is $\gamma_N\approx  0.035$, so
that  the maximal  DL length  is $\xi  \approx 28\approx  D/2$.  We  have
checked that the  approximated relation $\xi \approx D/2$  holds also for
other values of $k$.\\
 
We observe that  the minimal saturated value of  $\gamma_N\approx 2/D$ is
attained immediately  after the  decrease of  $\gamma_N$ in  the interval
$N\leq 3$.  This behavior was observed also for other values of $k$, with
$N$ in  the interval $N<k/2$.  The  decrease in $\gamma_N$ may  be due to
the fact that for $\vert r\vert  < k/2$ the hopping constants (\ref{bes})
are of the same order of magnitude, so that the effect of each additional
hopping constant is to increase the DL length.  For $\vert r\vert > k/2$,
on the  other hand, the  hopping constants (\ref{bes}) decay  faster than
exponentially, and their effect on  $\gamma_N$ turns out to be negligible
in the global-chaos  regime, as shown in  Fig. 7.  This fact  was used in
Ref. \cite{she} to justify the  truncation of the tight-binding model for
the KR at $N\approx k/2$.\\
 
We see  in Fig. 7,  however, that the  effect of the  exponentially small
hopping constants  for $\vert r\vert >  k/2$ is {\it not  negligible} for
small values of  $\tau < 1$, corresponding to local-chaos  regimes or the
vicinity of QAR.  In sharp contrast with the case of $N<k/2$, these small
hopping constants have  the somehow paradoxical effect  to {\it increase}
$\gamma_N$, i.e. to  {\it decrease} the localization  length $\xi$.  This
effect appears to continue up to  a maximal value $N^{\ast}$ of $N$.  For
$N>N^{\ast}$,    $\gamma_N$    saturates     to    a    value    $\approx
\gamma_{N^{\ast}}$.  Both $N^{\ast}$  and $\gamma_{N^{\ast}}$ increase as
$\tau$ or $k$ decrease.\\
 
While we are  unable to provide at this point  a quantitative explanation
of these phenomena, it will become apparent from the following discussion
that  they  are  quite  natural.   We already  know  that  the  truncated
tight-binding model is  exactly equivalent to a dynamical  problem with a
potential  $V_N(\theta )$  [see (\ref{VN})].   The region  of analyticity
${\cal R}_N$  of $V_N(\theta )$  increases with $N$ for  $N$ sufficiently
large ($N>k/2$).   The strip of  analyticity ${\cal R}_{QE,N}$ of  the QE
states has  a width $2\gamma_N$,  where $\gamma_N$ depends on  $\tau$ and
determines  the   rate  of  exponential   decay  of  the  QE   states  in
angular-momentum  space.  In  the infinitesimal  vicinity of  QAR, ${\cal
R}_{QE,N}={\cal  R}_N$, so  that an  increase in  the analyticity  of the
potential results in  a corresponding increase of the  analyticity of the
QE  states.  For  finite $\tau$,  however, the  increase of  ${\cal R}_N$
leads to  an increase of ${\cal  R}_{QE,N}$ only up to  $N\sim N^{\ast}$,
where  $\gamma_N$  saturates  to  a value  $\gamma_{N^{\ast}}$.   As  the
strength  of   chaos  is  increased   by  increasing  $\tau$   (or  $K$),
$\gamma_{N^{\ast}}$ decreases  (the DL  length $\xi$ increases).   At the
same time, $N^{\ast}$ decreases, so that the influence of the analyticity
of  $V_N(\theta )$  on  the analyticity  of the  QE  states is  gradually
reduced.  The extreme case corresponds  to the global-chaos regime.  Here
the increase  of ${\cal  R}_N$ does  not lead to  any increase  in ${\cal
R}_{QE,N}$,  and   $\gamma_{N^{\ast}}$  is  totally  determined   by  the
diffusion coefficient $D$.  This case of  DL in a semiclassical regime of
global  chaos  is  thus  completely  different in  nature  from  that  of
QAR-localization.  We may  now say that the transition  between these two
kinds of quantum localization takes place via {\it a gradual reduction of
the influence of the analyticity of the  system on that of the QE states,
as the level of chaos is increased}.\\
 
This characterization  of the  transition is  quite natural  as it  has a
well-known  classical analogue.   When  a  nonintegrability parameter  is
increased, the analyticity of functions representing classical structures
(e.g., KAM tori, periodic orbits)  generally decreases.  A famous example
is the  golden-mean torus in  the standard  map.  When the  parameter $K$
approaches  from below  the critical  value $K_c\approx  0.9716$ for  the
disappearance  of this  torus, the  width $\Delta_{GM}$  of the  strip of
analyticity  of  functions representing  the  torus  shrinks to  zero  as
$K_c-K$ (for $K$  sufficiently close to $K_c$)  \cite{shen}.  At $K=K_c$,
the torus  is a  continuous nondifferentiable curve,  and for  $K>K_c$ it
becomes a cantorus  allowing for global chaotic  diffusion.  The decrease
of $\Delta_{GM}$  as $K$  is increased  is analogous  to the  decrease of
$\gamma_{N^{\ast}}$.  Unlike $\Delta_{GM}$, however, the minimal value of
$\gamma_{N^{\ast}}$, in  the global-chaos  regime, is  not zero  but only
inversely proportional to $D\gg 1$.\\
 
According  to  this  picture,  a   saturation  of  $\gamma_N$  around  an
asymptotic value $\approx \gamma_{N^{\ast}}$ should occur for arbitrarily
low level of chaos (arbitrarily small  $\tau$ or $k$), but $N^{\ast}$ and
$\gamma_{N^{\ast}}$  may be  very  large.  We  thus  expect that,  unlike
QAR-localization, the DL  decay can never be faster  than exponential but
it should always feature an  asymptotic exponential tail.  For very small
$\tau$, this  asymptotic behavior may start  only at very small  value of
the wave function,  and the decay may look faster  than exponential above
this value.  This is probably what one sees in Fig. 4 for $\tau =10^{-5}$
and $\tau =10^{-3}$.  Here an exponential  tail may be observed only much
below  the  level  of  numerical noise  ($\sim  10^{-30}$).   Because  of
computational limitations,  we were not  able to verify the  existence of
the saturation effect and the associated exponential tail in the DL decay
for $\tau < 0.05$.\\
 
In Fig.  8, we  plot $\gamma_N$  for $N=12$  and $N=1$  as a  function of
$\tau$  near QAR.   While  $\gamma_{12}$ changes  significantly from  its
maximal  value  (near  QAR)  to  its  minimal  DL  value  $\approx  2/D$,
$\gamma_1$  appears to  be independent  of  $\tau$.  In  fact, the  small
fluctuations in  $\gamma_1$ are  of the  same order  of magnitude  as the
numerical error  in this quantity.   The case  of $N=1$ corresponds  to a
nearest-neighbor  pseudorandom  ``Lloyd  model"  \cite{fish,she,fis,gri}.
The independence of the localization length $\xi =1/\gamma$ on $\tau$ for
such a model has been verified numerically \cite{she,gri} for many values
of  a coupling  constant (e.g.,  $k$).  The  potential $V_1(\theta  )$ in
(\ref{VN}) is a special case of (\ref{atan}),
\begin{equation}\label{atan0}
V_1(\theta ) = -\frac{2}{k}\arctan [\kappa \cos (\theta)] \ ,
\end{equation}
where   $\kappa  =-2J_1(k/2)/J_0(k/2)$.    For  $A=-2/k$   and  arbitrary
$\kappa_0$, the potential (\ref{atan})  leads to nearest-neighbor hopping
constants  also in  the original  $M=2$ tight-binding  model (\ref{And}):
$W_n=0$  except  of  $W_{\pm  1}=\kappa   /2$.   This  is  a  two-channel
pseudorandom ``Lloyd  model" (see some  generalizations of this  model in
Appendix C).\\
 
In Ref.  \cite{fis} the relation $\xi  =D/2$ was derived for  the KR with
the potential  (\ref{atan0}) by  assuming that the  pseudorandom disorder
can  be replaced  by a  truly random  one, giving  the usual  Lloyd model
\cite{llo}.   The  validity  of  this assumption  was  verified  in  Ref.
\cite{gri}.   We now  give an  alternative and  simpler ``proof"  of this
relation, based  on the {\it  exact} result $\xi_0=D/2$ which  is derived
below.  The relation $\xi =D/2$ is an immediate consequence of the latter
result and the numerically observed independence of $\gamma_1$ on $\tau$.
The derivation  below of $\xi_0=D/2$  makes use,  of course, of  the fact
that  the KR  with the  potential (\ref{atan0})  exhibits QAR  (of period
two).  As shown  in Sec.  II, this  QAR can occur only  if $V(\theta +\pi
)+V(\theta  )=\mbox{constant}$.   This  condition  is  satisfied  by  the
potential  (\ref{atan}) only  if  $\kappa_0=0$ with  $\mbox{constant}=0$,
which  is precisely  the  case  of (\ref{atan0}).   As  shown below,  the
relation $\xi =D/2$ is valid also  for the equivalent system of the $M=2$
MKR with (\ref{atan0}).\\
 
To  derive the  relation  $\xi_0=D/2$, we  first  perform explicitly  the
integral in (\ref{D0}) for the potential (\ref{atan0}).  The final result
can be easily expressed in terms of $\gamma$ in (\ref{gamma}):
\begin{equation}\label{Dgamma}
D = 2\langle c^2\rangle \left [ \frac{1}{\gamma} +
    \frac{1}{2} + O(\gamma ^2)\right ]\ ,
\end{equation}
where  we  have  assumed  the  strong-chaos regime  $\kappa  \gg  1$  for
(\ref{atan0}),  corresponding to  very small  $\gamma$ in  (\ref{gamma}).
Now,   the  QAR-localization   length   $\xi_0$   for  (\ref{atan0})   is
$\xi_0=1/\gamma$ (see Sec.  IV), and $\langle c^2\rangle =1$ for both the
KR and the $M=2$ MKR.  The relation $\xi_0=D/2$ for these systems follows
then from Eq. (\ref{Dgamma}) in the strong-chaos regime.\\
 
From Fig. 7 it appears that  $\gamma_2$ is also independent of $\tau$ for
the particular value  of $k$ chosen.  The case of  $N=2$ corresponds to a
next nearest-neighbor pseudorandom model.   Such a model was investigated
by  Brenner  and  Fishman  \cite{gri} who  found  that  its  localization
properties  are   quite  different  from   those  of  its   truly  random
counterpart.  In particular, the  localization length strongly depends on
the  irrational number  (analogous to  $\eta =\tau  /2\pi$) defining  the
pseudorandom disorder.  Fig. 7 suggests, however, that it may be possible
to define  pseudorandom tight-binding models with  $N>1$ neighbors, whose
localization length is almost independent  of $\tau$.  Such models should
exhibit localization  properties quite  similar to  those of  their truly
random counterparts.  The existence of  these models and related problems
will be investigated in a future work \cite{next}.\\
 
\newpage
 
\begin{center} {\bf VII. CONCLUSIONS}\\ \end{center}
 
In this  paper, the  problem of  quantum localization  in 1.5  degrees of
freedom (``minimal chaos")  has been studied in a large  class of systems
exhibiting  classical  chaotic  diffusion, the  modulated  kicked  rotors
(MKRs).   These   systems  feature  two  basically   different  kinds  of
asymptotic  exponential  localization   in  angular-momentum  space:  (a)
Dynamical localization  (DL), which, as in  the case of the  KR and other
systems  \cite{fish,she,blu,kr2,kr3}, is  the  localization exhibited  by
pseudorandom  tight-binding models  for  irrational values  of $\eta$  (a
scaled   $\hbar$)  \cite{not}.   (b)  QAR-localization,   which  is   the
localization occurring for $\eta$ in the infinitesimal vicinity of points
$\eta_0$  where  the  quantum   dynamics  is  exactly  periodic  [quantum
antiresonance (QAR)].\\
 
The existence of QAR-localization has  been rigorously established in the
framework  of   a  self-consistent  approach  for   both  period-one  and
period-two  QAR (see  Sec.  III  and Appendix  A; it  seems that  QARs of
period  larger than  two  do  not exist  for  nonintegrable MKRs).   This
approach  leads  to the  basic  equation  (\ref{eq}) [or  (\ref{geq})  in
Appendix A] for the QE problem  in the infinitesimal vicinity of QAR.  It
follows  from this  equation  that  the QE  spectrum  is  pure point  for
infinitesimal $\eta  -\eta_0$, and  that the  asymptotic QAR-localization
length $\xi_0$ is completely determined from the analytical properties of
the potential  appearing in  the equation:  $\xi_0 \leq  1/\gamma$, where
$\gamma$ is the smallest distance of  a singularity of the potential from
the  real  $\theta$-axis.   In  many   interesting  cases,  such  as  the
potentials (\ref{atan}) and (\ref{VN}), $\xi_0=1/\gamma$ exactly.\\
 
Being associated  with the  1D time-independent  Schr\"{o}dinger equation
(\ref{eq}), QAR-localization is of an  ``integrable" nature.  In fact, if
$\eta_0=0$ is a QAR point, the Hamiltonian in (\ref{eq}) is precisely the
integrable limit  ($T\rightarrow 0$  at fixed $k$)  of the  classical MKR
Hamiltonian (\ref{MKR}).   On the other  hand, DL is associated  with the
difference equations  (\ref{tbm}) [or  (\ref{And})] for  the pseudorandom
tight-binding   models.   These   equations  depend   not  only   on  the
dimensionless kicking  parameter $k$  [essentially the only  parameter in
(\ref{eq})],  but  also  on  $\eta$, which  determines  the  pseudorandom
disorder.  This pseudorandomness, which is absent in the angular-momentum
representation  of (\ref{eq}),  introduces  ``nonintegrable" features  in
DL.\\
 
Thus, while the asymptotic  QAR-localization length $\xi_0$ is completely
determined by the  analytical properties of the  potential in (\ref{eq}),
the  asymptotic DL  length $\xi$  depends, in  general, also  on $k$  and
$\eta$.   Starting  from  the   integrable  case  $\eta  =\eta_0=0$,  and
increasing the  level of  classical chaos by  increasing $\eta$  at fixed
$k\gg 1$,  we find that  the sensitivity of $\xi$  to an increase  in the
analyticity of the  potential (exhibited by the hopping  constants in the
tight-binding model)  is gradually  reduced.  This phenomenon,  which has
been considered in detail in Sec.  VI  and is clearly shown in Fig. 7, is
a vivid manifestation of classical chaos in quantum dynamics.  As soon as
one  reaches  a  semiclassical  regime  ($\eta  <1$)  of  global  chaotic
diffusion ($K=4\pi \eta  k >1$), $\xi$ becomes totally  unaffected by any
increase in the  analyticity of the potential.  In this  regime, $\xi$ is
completely determined by the classical chaotic-diffusion coefficient $D$,
$\xi \approx D/2$.\\
 
While   QAR-localization   is   basically  different   from   DL,   these
localizations  are   expected  to  look   quite  similar  if   $\eta$  is
sufficiently  irrational and  close to  $\eta_0$.  The  numerical results
shown in  Figs. 2-7  strongly support  this expectation.   In particular,
Figs. 5-7 show that in cases where the QAR-localization length $\xi_0$ is
finite  ($\xi_0\neq 0$)  the  DL  length $\xi$  is  well approximated  by
$\xi_0$ if $\eta$ is sufficiently close  to $\eta_0$.  In the case of the
potential  (\ref{atan0}), corresponding  to a  pseudorandom Lloyd  model,
$\xi$ appears to  be independent of $\eta$, so that  $\xi =\xi_0$.  Using
then the fact that the KR  with the potential (\ref{atan0}) exhibits QAR,
and $\xi_0=D/2$ exactly, the relation  $\xi =D/2$ for this system follows
immediately.   Ref. \cite{fis}  presents a  much lengthier  derivation of
this relation, based on the assumption that the pseudorandom disorder can
be replaced by a truly random one.\\
 
In summary, the presence of QAR  in nonintegrable systems is quite useful
for  studying  several  interesting  aspects of  DL,  in  particular  the
transition from DL  in local-chaos or almost-integrability  regimes to DL
in the global-chaos regime, and  for obtaining exact lower bounds $\xi_0$
to the asymptotic DL length $\xi$.\\
 
 
{\bf Acknowledgments}\\
 
We would like  to thank M.  Feingold, S.  Fishman,  U.  Smilansky, and F.
M.   Izrailev for  useful discussions  and comments.   We are  especially
grateful  to M.   Feingold for  providing us  with an  efficient computer
program for the calculation of the Lyapunov spectra of products of random
matrices.  This  work was partially  supported by the Israel  Ministry of
Science and Technology and the  Israel Science Foundation administered by
the Israel Academy of Sciences and Humanities.
 
\newpage
 
\appendix{\bf GENERAL QAR OF PERIOD ONE AND PERIOD TWO\\ IN NONINTEGRABLE
MKRs}{\ }\\
 
In Sec.  III, we have restricted our attention, for simplicity, to a case
of period one  QAR occurring when $\tau_j$ in (\ref{tauj})  is a multiple
of $2\pi$,  $\tau_j=2\pi m_j$.  Here  we shall consider the  more general
case of  $\tau_j$ equal to  a multiple of $\pi$,  $\tau_j=m_j\pi$, giving
both  period-one  and  period-two  QARs.   We  believe,  but  are  unable
presently to  give an exact  proof, that this  is actually the  {\it most
general} case  of QAR in  the nonintegrable MKRs (\ref{MKR}).   We derive
the QAR effective  Hamiltonian [analogous to (\ref{G1})]  in this general
case,  and  obtain  an  interesting relation  [Rel.  (\ref{ke12})  below]
between the values of $k_{eff}$ associated with period-one and period-two
QARs in the $M=3$ MKR.\\
 
By  repeated   application  of   Rel.  (\ref{pi}),   we  find   that  for
$\tau_j=m_j\pi$ the  evolution operator (\ref{UMKR}) can  be expressed as
follows
\begin{equation}\label{U2MKR}
U = \exp\left [ -ik\sum_{j=0}^{M-1} c_jV(\theta +{\bar m}_j\pi )\right ]
    \exp (-i{\bar m}_0\pi {\hat n}^2)\ ,
\end{equation}
where ${\bar m}_j=\sum_{s=j}^{M-1} m_s$.   Consider first period-one QAR.
It is  clear from  (\ref{U2MKR}) that  this QAR may  be possible  only if
${\bar   m_0}$    is   even.    In    this   case,   one    has   $U=\exp
[-ik\overline{V}(\theta  )]$,   where  $\overline{V}(\theta  )$   is  the
function having Fourier coefficients
\begin{equation}\label{v1bar}
{\bar v}_n = v_n\sum_{j=0}^{M-1} c_j e^{i n{\bar m}_j}\ .
\end{equation}
Here $v_n$ are the Fourier coefficients of $V(\theta )$.  We denote $c_j$
by $c_{e,j}$ or  $c_{o,j}$ depending on whether  the corresponding ${\bar
m}_j$ is even  or odd, respectively.  The sum over  all $c_{e,j}$ (or all
$c_{o,j}$) will be denoted by $c_e=\sum c_{e,j}$ (or $c_o=\sum c_{o,j}$).
Rel. (\ref{v1bar}) can then be written as follows
\begin{equation}\label{v1bare}
{\bar v}_n = v_n [c_e+(-1)^nc_o]\ .
\end{equation}
Now, period-one QAR,  i.e., $U=1$ [without loss of  generality, we assume
that $\beta =0$ in (\ref{QAR})], implies that ${\bar v}_n=0$ for all $n$.
From  Rel. (\ref{v1bare})  we then  obtain the  following conditions  for
period-one QAR, corresponding to two different cases:
\begin{equation}\label{s10}
\begin{array}{ll}
c_e - c_o = 0 & \mbox{\ \ \ \ \ if $V(\theta +\pi )=-V(\theta )$,} \\
c_e = c_o = 0 & \mbox{\ \ \ \ \ otherwise.}
\end{array}
\end{equation}
Notice that the second condition [for general $V(\theta )$]  leads to the
trivial result $c_j=0$ for all $j$ if each of the sets $\{ c_{e,j}\}$ and
$\{ c_{o,j}\}$ contains {\it only one}  element.  This may happen only if
$M=2$  and $m_0$  and $m_1$  are  both odd.   Conditions (\ref{s10})  for
$\tau_j=m_j\pi$  are a  generalization  of condition  (\ref{s0}) in  Sec.
III.\\
 
If ${\bar m}_0$ is odd, we show that period-two QAR takes place.  We find
in  this case  that  $U^2=\exp  [-ik\overline{V}^{(2)}(\theta )]$,  where
$\overline{V}^{(2)}(\theta )$ has Fourier coefficients
\begin{equation}\label{v2bar}
{\bar v}^{(2)}_n = v_n\sum_{j=0}^{2M-1} c_j e^{i n{\bar m}'_j}\ .
\end{equation}
Here  ${\bar   m}'_j=\sum_{s=j}^{2M-1}m_s$,  and  $c_j$  and   $m_j$  are
``extended" beyond $j=M-1$ by defining,  for $j\geq M$, $c_j=c_{j-M}$ and
$m_j=m_{j-M}$.  From the  definitions of ${\bar m}'_j$  and ${\bar m}_j$,
we see that ${\bar m}'_j={\bar m}_{j-M}-{\bar m}_0$ for $j\geq M$.  Using
the   last    relation   in   (\ref{v2bar}),   we    find   that   ${\bar
v}^{(2)}_n=2v_n\sum_{j=0}^{M-1}c_j$ for $n$ even and ${\bar v}^{(2)}_n=0$
for $n$ odd.  The conditions for  period-two QAR, i.e., $U^2=1$ or ${\bar
v}^{(2)}_n=0$ for all $n$, are therefore
\begin{equation}\label{s20}
\begin{array}{ll}
c_j \mbox{\ \ arbitrary} &
\mbox{\ \ \ \ \ if $V(\theta +\pi )=-V(\theta )$,} \\
\sum_{j=0}^{M-1}c_j=0    & \mbox{\ \ \ \ \ otherwise.}
\end{array}
\end{equation}
Thus, if $V(\theta  +\pi )=-V(\theta )$ and ${\bar m}_0$  is odd, one has
period-two  QAR  for  {\it  arbitrary}   values  of  $c_j$.   This  is  a
considerable generalization  of the  period-two QAR in  the KR  (see Sec.
II),  discovered by  Izrailev  and Shepelyansky  \cite{izr}.  The  second
condition  in (\ref{s20})  [for  $V(\theta +\pi  )\neq  -V(\theta )$]  is
precisely condition (\ref{s0}) in Sec.  III.\\
 
We now  consider small  perturbations of $\tau_j$  near their  QAR values
$m_j\pi$.  For definiteness, we shall work out in detail here the case of
period-one QAR, i.e., ${\bar m}_0$ even, but we shall show at the end how
to  extend  the  results   to  period-two  QAR.   Writing  $\tau_j=m_j\pi
+\epsilon_j$  in  (\ref{UMKR}),  and  formally  expanding  in  powers  of
$\epsilon_j$ as in Sec.III, we find, to first order in $\epsilon_j$,
\begin{equation}\label{gU1}
U\approx 1 - \sum_{j=0}^{M-1}\epsilon_j
\left \{ i{\hat n}^2 - k [2iV_j'(\theta ) {\hat n}+V_j''(\theta )] +
ik^2 V_j^{\prime 2}(\theta )\right \} \ ,
\end{equation}
where  $V_j(\theta )=\sum_{s=0}^jc_sV(\theta  +{\bar  m}_s\pi )$.   Using
(\ref{epj}), the expression in (\ref{gU1}) can be written, to first order
in $\epsilon$, as $\exp (-i\epsilon G_1)$, where
\begin{equation}\label{gG1}
G_1 = \left [ {\hat n} -kV_a'(\theta )\right ]^2 +
k^2 [\Delta V'(\theta )]^2 \ .
\end{equation}
Here
\begin{equation}\label{adV}
V_a(\theta )=\sum_{j=0}^{M-1}\frac{t_{j+1}-t_j}{T} V_j(\theta )\ ,
\ \ \ \ \ [\Delta V'(\theta )]^2=\sum_{j=0}^{M-1}\frac{t_{j+1}-t_j}{T}
V_j^{\prime 2}(\theta )-V_a^{\prime 2}(\theta )\ .
\end{equation}
The QE problem in the infinitesimal  vicinity of QAR is $G_1\psi =g\psi$,
and after the gauge transformation
\begin{equation}\label{gga}
\varphi = \exp \left [ -ikV_a(\theta )\right ] \psi \ ,
\end{equation}
it reduces to the 1D Schr\"{o}dinger equation
\begin{equation}\label{geq}
-\frac{d^2\varphi}{d\theta ^2} +
k^2[\Delta V'(\theta )]^2\varphi = g\varphi \ .
\end{equation}
The analysis  of Eq.  (\ref{geq}) is  completely similar  to that  of Eq.
(\ref{eq})  in  Sec.   III,  and   the  conclusions  are  the  same:  the
QAR-localization  length   $\xi_0$  in   entirely  determined   from  the
analytical properties of the function $[\Delta V'(\theta )]^2$.\\
 
The results  (\ref{gG1}) and (\ref{geq})  are valid  also in the  case of
period-two  QAR (${\bar  m}_0$  odd),  but all  the  sums  over $j$  [see
(\ref{gU1})   and   (\ref{adV})]  run   now   from   $j=0$  to   $j=2M-1$
($t_j=t_{j-M}$  for  $j\geq  M$),  and   $V_j(\theta  )$  is  defined  as
$V_j(\theta )=\sum_{s=0}^jc_sV(\theta +{\bar m}'_s\pi )$.\\
 
As an interesting example, we consider  the $M=3$ MKR with $V(\theta +\pi
)=-V(\theta )$  and $m_j$ independent  of $j=0,\ 1,\ 2$.   For $m_j=2$,
one has the simple case of period-one QAR treated in Sec.  III, provided,
of course, $c_0+c_1+c_2=0$.  The value  of $k_{eff}$ in (\ref{G1}) can be
easily expressed in terms of  two independent coefficients, say $c_0$ and
$c_1$, using (\ref{avb}) [$(t_{j+1}-t_j)/T=1/3$ for all $j$]:
\begin{equation}\label{ke1}
k_{eff}=k[2(c_0^2+c_1^2+c_0c_1)/9]^{1/2}\ .
\end{equation}
 
Suppose now  that the period  $T$ is halved, $T\rightarrow  T/2$, leaving
all other parameters  (including $c_j$) unchanged.  Then  $m_j=1$ for all
$j$, and,  since ${\bar m}_0=3$  (odd) one  now has period-two  QAR.  The
``real-time" period  remains then  $T$.  Notice  also that  the condition
$c_0+c_1+c_2=0$  is  consistent with  (\ref{s20})  [$c_j$  can be  chosen
arbitrarily  in this  case!].   Using $V(\theta  +\pi  )=-V(\theta )$  in
(\ref{adV})     [with      $M\rightarrow     2M$      and     $V_j(\theta
)=\sum_{s=0}^jc_sV(\theta +{\bar m}'_s\pi )$], we get
\[
k^2 [\Delta V'(\theta )]^2 = [k_{eff}^{(2)}]^2 V^{\prime 2}(\theta )\ ,
\]
where
\begin{equation}\label{ke2}
k_{eff}^{(2)}=k[5(c_0^2+c_1^2+c_0c_1)/6]^{1/2}\ .
\end{equation}
By comparing (\ref{ke2}) with (\ref{ke1}), we obtain the simple relation,
valid for all values of $c_0$ and $c_1$,
\begin{equation}\label{ke12}
\frac{k_{eff}^{(2)}}{k_{eff}}=(15/4)^{1/2}\ .
\end{equation}
We  were not  able to  discover similar  simple relations  for other  MKR
systems.\\
 
\appendix{\bf GENERAL QAR IN INTEGRABLE MKR SYSTEMS}{\ }\\
 
It is instructive  to study the QAR phenomenon in  a ``linear" version of
the MKR, defined by the general Hamiltonian:
\begin{equation}\label{LMKR}
H=\frac{\tau}{T}L+{\hat k}V(\theta )\sum_{j=0}^{M-1}c_j\Delta_T(t-t_j)\ .
\end{equation}
We show that  the system (\ref{LMKR}) is equivalent,  effectively, to the
linear  KR  (\ref{LKR}), which  is  integrable  \cite{berry} and  exactly
solvable to a large extent \cite{exa}.  We then derive general conditions
for QAR of  arbitrary period $p$ in (\ref{QAR}), and  show rigorously the
existence of  exponential localization  in the infinitesimal  vicinity of
QAR.\\
 
The evolution operator  for (\ref{LMKR}), from $t=-0$ to  $t=T-0$, can be
expressed as follows
\begin{equation}\label{ULMKR}
U = \prod_{j=0}^{M-1}\exp (-i\tau_j{\hat n})\exp [-ic_jkV(\theta )] =
    \exp (-i\tau {\hat n}) \exp [-ik\overline{V}(\theta )]\ ,
\end{equation}
where $\tau_j = (t_{j+1}-t_j)\tau /T$,
\begin{equation}\label{Vbar}
\overline{V}(\theta ) = \sum_{j=0}^{M-1}c_jV(\theta +\chi_j )]\ ,
\end{equation}
$\chi_j=\sum_{s=0}^{j-1}\tau_j$  for  $j\geq   1$,  and  $\chi_0=0$.   By
comparing  the  last  expression  for   $U$  in  (\ref{ULMKR})  with  Eq.
(\ref{ULKR}), we see  that the problem has been  reduced, essentially, to
that  of  the  linear  KR   \cite{exa}  with  an  ``effective"  potential
$\overline{V}(\theta )$.   Thus, from the  discussion at the end  of Sec.
II, it  follows that QAR  with arbitrary  period $p$ occurs  precisely at
rational values  of $\eta =\tau  /2\pi =m/p$, provided  ${\bar v}_{sp}=0$
for  all  $s$.   Here  ${\bar  v}_n$  are  the  Fourier  coefficients  of
$\overline{V}(\theta )$ in (\ref{Vbar}), and are given by
\begin{equation}\label{vbar}
{\bar v}_n = v_n\sum_{j=0}^{M-1} c_j e^{i n\chi_j}\ ,
\end{equation}
where $v_n$  are the  Fourier coefficients  of $V(\theta  )$.  It  is now
clear from  (\ref{vbar}) that, in contrast  with the linear KR  case (see
Sec.  II), the  requirement ${\bar v}_{sp}=0$ does  not necessarily imply
the vanishing of $v_{sp}$.  For simplicity, let us assume in what follows
that all  $\tau_j$'s are  equal, $\tau_j=\tau  /M$ for  all $j$,  so that
$\chi_j=j\tau /M$.   Then, with  $\tau =2\pi  m/p$, the  condition ${\bar
v}_{sp}=0$ is satisfied for all $s$ if
\begin{equation}\label{cbar}
{\bar c}_s \equiv \sum_{j=0}^{M-1} c_j \exp (2\pi ijsm/M) = 0\ .
\end{equation}
Now, if  $m$ and $M$ are  relatively prime, the sequence  ${\bar c}_s$ in
(\ref{cbar})  is, up  to some  rearrangement for  $m\neq 1$,  the Fourier
transform  of   $c_j$  ($j,\   s=0,...,\  M-1$).   Then   ${\bar  c}_s=0$
necessarily  implies that  $c_j=0$, and  the QAR  is trivial  for generic
potentials  $V(\theta )$.   If, on  the other  hand, $m$  and $M$  have a
maximal  common factor  $\overline{m}>1$,  the  condition ${\bar  c}_s=0$
implies only that
\begin{equation}\label{cqarl}
\sum_{r=0}^{\overline{m}-1} c_{j+rM'} = 0\ ,
\end{equation}
for $j=0,...,\ M'-1$, where  $M'=M/\overline{m}$.  To summarize, the $M'$
equations (\ref{cqarl})  are the necessary and  sufficient conditions for
QAR of period $p$ in the  linear MKRs with equal $\tau_j$'s ($\tau_j=2\pi
m/pM$) and arbitrary  potentials $V(\theta )$.  These  conditions lead to
nontrivial results only if $\overline{m} > 1$.  It should be noticed that
the conditions  (\ref{cqarl}) do not  depend on  $p$, i.e., they  are the
same as those for the fundamental ($p=1$) QAR with $\tau =2\pi m$.\\
 
We  now  consider  small  perturbations of  $\tau$  near  the  conditions
(\ref{cqarl}).  Since the evolution operator (\ref{ULMKR}) is the same as
that of a linear KR  with effective potential $\overline{V}(\theta )$, an
exact  expression  of  the  QE   states  for  irrational  $\eta$  can  be
immediately written using the results of Ref. \cite{exa}:
\begin{equation}\label{qel}
\psi_l(\theta ) = \exp [i\varphi_l(\theta )]\ ,
\end{equation}
where, for all  integers $l$, $\varphi_l(\theta )=l\theta  + \phi (\theta
)$, and $\phi (\theta )$ is a periodic function with Fourier coefficients
\begin{equation}\label{phibar}
{\bar \phi}_n =
\frac{k{\bar v}_n}{1-e^{in\tau}}\ \ \ (n\neq 0)\ ,
\end{equation}
${\bar v}_n$  being the Fourier coefficients  of $\overline{V}(\theta )$,
and ${\bar \phi}_0$ is arbitrary.  The corresponding quasienergy is given
by  $\omega_l={\bar v}_0  + l\tau$.   Now, in  the ordinary  case of  the
linear  KR, the  result analogous  to (\ref{phibar})  [with ${\bar  v}_n$
replaced by the given coefficients $v_n$  of $V(\theta )$] is clearly not
defined for $n=sp$  and $\eta =m/p$, i.e., at quantum  resonance.  In our
case, however, with the coefficients  ${\bar v}_n$ given by (\ref{vbar}),
it is easy to show that  the expression (\ref{phibar}) for $n=sp$ is well
defined in the QAR limit $\eta \rightarrow m/p$, the limit being taken on
some  sequence of  irrational  $\eta$'s converging  to  $m/p$.  In  fact,
writing   $\tau  =(2\pi   m+\epsilon  )/p$   and  using   the  conditions
(\ref{cqarl}), we find from (\ref{vbar}) (with $\chi_j=j\tau /M$) that
\begin{equation}\label{vbare}
{\bar v}_{sp} = v_{sp}\sum_{j=0}^{M'-1} e^{2\pi ijsm'/M'}
\sum_{r=0}^{\overline{m}-1} c_{j+rM'}
\left ( e^{i s(j+rM')\epsilon /M} - 1\right ) \ .
\end{equation}
After   substituting   (\ref{vbare})   into  (\ref{phibar})   and   using
(\ref{cqarl}), we obtain
\begin{equation}\label{phibare}
\lim_{\eta \rightarrow m/p} {\bar \phi}_{sp} =
-\frac{kv_{sp}}{\overline{m}}\sum_{j=0}^{M'-1} e^{2\pi ijsm'/M'}
\sum_{r=0}^{\overline{m}-1} r c_{j+rM'}\ .
\end{equation}
Rel. (\ref{phibare})  shows that the  QE states  are well defined  in the
infinitesimal vicinity of  the QAR of arbitrary period $p$.   We see from
(\ref{phibar}) and (\ref{phibare}) that these  QE states are localized in
$L$ space with  the same localization length of  the (analytic) potential
$V(\theta )$.  In  accordance with this, the QE spectrum  of $U^p$ in the
limit  of infinitesimal  $\epsilon$  is  pure-point, $\omega_l\bmod  2\pi
=\epsilon l$.\\
 
These rigorous  results, which have no  analogue in the ordinary  case of
the  linear KR  \cite{exa}, can  be derived  by an  alternative approach,
similar to that used for the  nonintegrable MKRs in Sec.  III.  The basic
evolution operator $U^p$ can be  expressed in the form (\ref{ULKRp}) with
$V(\theta )$ replaced by $\overline{V}(\theta )$.  Then, for $p\tau =2\pi
m+\epsilon$, one easily finds, using  (\ref{cqarl}), that $U^p$ is given,
to first order in $\epsilon$, by $\exp (-i\epsilon G_1)$, where
\begin{equation}\label{LG1}
G_1 = {\hat n} - \phi_0^{\prime }(\theta )\ ,
\end{equation}
with $\phi_0(\theta  )=\lim_{\epsilon \rightarrow 0}\phi (\theta  )$.  In
the  case of  the  fundamental  QAR ($p=1$),  with  $\tau_j=2\pi m_j$  in
(\ref{ULMKR})  as in  Sec.   III, we  find  that $\phi_0(\theta  )=k{\bar
d}V(\theta )$.   Thus the operator  (\ref{LG1}) may be considered  as the
linear  version  of  (\ref{G1}).   Its eigenvalues  are  simply  all  the
integers $l$,  and its  eigenstates are given  by (\ref{qel})  with $\phi
(\theta)$ replaced by  $\phi_0(\theta )$.  We then see that  in the limit
$\epsilon  \rightarrow  0$  the  QE   spectrum  and  eigenstates  of  the
approximate  evolution  operator $\overline{U}^p=\exp  (-i\epsilon  G_1)$
agree  precisely with  the rigorous  ones  obtained above.   This may  be
evidence that  the results  obtained by  the self-consistent  approach in
Sec.  III are, in fact, rigorous.\\
 
\appendix{\bf  NEARLY-SINGULAR  CASES  OF\\  MULTI-CHANNEL  TIGHT-BINDING
MODELS}{\ }\\
 
Clearly, all  the hopping  constants $\overline{W}_{s,n}$, $s\neq  0$, in
the singular  case [Eq. (\ref{sinW})] have  the same range in  $n$ (i.e.,
the range of $W_n$).  It  is thus interesting to consider nearly-singular
cases   of  (\ref{tbm})   ($M>2$)   for  which   the  hopping   constants
$\overline{W}_{0,n}$    within    a    channel   are    nonzero,    while
$\overline{W}_{s,n}$  for $s\neq  0$  still have  approximately the  same
range in $n$.   The simplest possible case where this  may happen is when
$c_j$ takes  only the  values $0,\  \pm 1,\  \pm 2$.   In this  case, the
Fourier  transform  $\overline{W}_s(\theta   )$  of  $\overline{W}_{s,n}$
assumes the relatively simple form
\begin{equation}\label{nsinW}
\overline{W}_s(\theta ) = c^{(1)}(s)W(\theta ) +
                          c^{(2)}(s)\frac{W(\theta )}{1-W^2(\theta )}\ ,
\end{equation}
where $c^{(1)}(s)$ and $c^{(2)}(s)$  are, respectively, the contributions
of the terms with $c_j=\pm 1$  and $c_j=\pm 2$ to $c(s)$ in (\ref{sinW}).
Using $c^{(1)}(0)+c^{(2)}(0)=0$ [Rel. (\ref{s0})], we find that
\begin{equation}\label{nsinW0}
\overline{W}_0(\theta ) =
-c^{(1)}(0)\frac{W^3(\theta )}{1-W^2(\theta )}\ .
\end{equation}
As  an example,  consider the  case  of $W(\theta  )=\kappa \cos  (\theta
)-\kappa_0$, which  corresponds to the potential  (\ref{atan}) ($A=-2/k$)
and  gives a  nearest-neighbor  (``Lloyd") model  (\ref{And}) for  $M=2$,
i.e., $W_n=0$ except of $W_{\pm 1}=\kappa /2$ (such a model is studied in
Sec.  VI).  We  see from (\ref{nsinW}) and (\ref{nsinW0})  that there are
two interesting  limits.  If  $\kappa_0\gg \max (\kappa  ,\ 1)$,  all the
hopping    constants    ${\overline     W}_{s,n}$    are    approximately
nearest-neighbor in  $n$, and  $E=c^{(1)}(0)\kappa_0$.  If, on  the other
hand,  $\max  (\kappa  ,\  \kappa_0)\ll  1$,  ${\overline  W}_{s,n}$  are
approximately nearest-neighbor for $s\neq 0$, while ${\overline W}_{0,n}$
are     approximately     next-next-nearest-neighbor    and     $E\approx
c^{(1)}(0)\kappa_0^3$.  The localization length $\xi =1/\gamma$, however,
depends only  on $W(\theta  )$, and  it is always  given exactly  by Rel.
(\ref{gamma}).\\
 
It should  be noticed, however,  that the choice $W(\theta  )=\kappa \cos
(\theta  )-\kappa_0$   is  not  a   good  one  for  obtaining   a  nearly
nearest-neighbor model  in the strong-chaos  regime ($D\gg 1$).   This is
because Eqs.  (\ref{Dgamma}) and  (\ref{gamma}) imply that  $\vert \kappa
\vert  ,\ \vert  \kappa /\kappa_0\vert  \gg 1$  in this  regime, so  that
$W(\theta )=\kappa \cos (\theta  )-\kappa_0=1$ for some $\theta$, leading
to  a  singularity in  Eqs.  (\ref{nsinW})  and (\ref{nsinW0}).   Let  us
choose, instead,
\begin{equation}\label{WM}
\frac{2W(\theta )}{1-W^2(\theta )}=-\tan [kV(\theta )]=
\kappa \cos (\theta )-\kappa_0 \ ,
\end{equation}
i.e.,
\begin{equation}\label{WMe}
W(\theta )=\frac{\kappa \cos (\theta )-\kappa_0}
{1+\{ 1+[\kappa \cos (\theta )-\kappa_0]^2\} ^{1/2}} \ .
\end{equation}
It  is clear  from (\ref{WMe})  that  in the  strong-chaos regime  $\vert
W(\theta )\vert \approx 1$ for  almost all $\theta$.  Eqs. (\ref{nsinW}),
(\ref{nsinW0}),  and (\ref{WM})  imply  then that  the hopping  constants
$\overline{W}_{s,n}$ are  all nearest-neighbor  in $n$ to  high accuracy,
i.e.,  $\overline{W}_{s,n}\approx   0$  except   of  $\overline{W}_{s,\pm
1}\approx  c^{(2)}(s)\kappa /4$.   The corresponding  tight-binding model
(\ref{tbm}) may thus be naturally  viewed as a multi-channel pseudorandom
Lloyd model for the strong-chaos regime.\\

\figure{Chaotic orbits  generated by  30,000 iterations of  the classical
Poincar\'{e} map  (\ref{map}) for  a $M=3$ Hamiltonian  (\ref{MKR}), with
$V(\theta )=\cos  (\theta )$,  $c_0=c_1=1$, $c_2=-2$, and  $t_{j+1}-t_j =
T/3$ for  all $j$: (a) Local  chaos for ${\hat k}=0.3$.  (b) Global chaos
for ${\hat k}=0.5$.\label{logl}}
\figure{Expectation value of the kinetic  energy, $E_0$, as a function of
the ``real  time" $t=\epsilon  s$, for  the MKR  system described  in the
caption  of  Fig.\  \ref{logl}  with  $k=2$ and  for  several  values  of
$\epsilon  =2\pi  (\eta  -\eta_0)$  ($\eta_0$   is  a  QAR  point).   The
continuous curve corresponds  to $\epsilon =0.01$, the  filled circles to
$\epsilon =0.11$, the squares to $\epsilon =0.17$, the filled diamonds to
$\epsilon =0.25$, and the triangles to $\epsilon =0.28$.\label{coll}}
\figure{Fourier transform $E_0(\nu  )$ of $E_0(t)$ (see  caption of Fig.\
\ref{coll}) for $\epsilon  =0.1$ and several values of  $k$ (see legend).
The symbols at  the bottom are the theoretical predictions  for the peaks
positions, based on  the eigenvalues of the  Mathieu equation (\ref{ma}).
The peaks  for $k=0.01$  and $k=1.0$  have been rescaled  by a  factor of
100,000 and 10, respectively, for visibility.\label{freq}}
\figure{Steady-state  probability distributions  $\langle \vert  \psi (n)
\vert  ^2\rangle$  over  angular  momentum $n\hbar$  for  the  $M=2$  MKR
($c_0=-c_1=1$ and $t_1=T/2$) with $V(\theta )=\cos (\theta )$ and $k=10$.
The three  curves correspond  to three irrational  values of  $\eta =\tau
/2\pi$: $\tau =10^{-5}$ (solid line),  $\tau =10^{-3}$ (dashed line), and
$\tau =10^{-1}$ (dotted line).  The saturation of $\langle \vert \psi (n)
\vert    ^2\rangle$   around    $10^{-30}$    is    due   to    numerical
noise.\label{fig4}}
\figure{Same as in Fig. 4, but for the potential (\ref{atan}) with $A=1$,
$\kappa =1$, $\kappa_0=0$, and $k=2$.  The straight thick dotted line has
slope $-2\gamma$,  where $\gamma$  is determined from  Eq. (\ref{gamma}).
For $\tau$ sufficiently small, this slope is expected to be very close to
the asymptotic rate of exponential decay of $\langle \vert \psi (n) \vert
^2\rangle$.  We  observe that this is  the case even for  $\tau$ not very
small, $\tau =10^{-1}$  (dotted line).  In fact,  this system corresponds
precisely to a  pseudorandom Lloyd model, whose DL length  $\xi$ seems to
be independent of $\tau$ (see Sec.  VI and Fig. 8).\label{fig5}}
\figure{Same as in Fig. 5, but for $k=10$.  The four curves correspond to
four  irrational values  of $\eta  =\tau /2\pi$:  $\tau =10^{-7}$  (solid
line), $\tau =10^{-5}$ (dashed  line), $\tau =10^{-3}$ (dot-dashed line),
and $\tau  =10^{-1}$ (dotted line).   The straight thick dotted  line has
slope $-2\gamma$,  where $\gamma$  is determined from  Eq. (\ref{gamma}).
\label{fig6}}
\figure{Minimal   Lyapunov  exponent   $\gamma_N$  associated   with  the
$N$th-order truncation  of the  tight-binding model  (\ref{tb2KR}) (i.e.,
$r$ restricted  to $\vert  r\vert \leq  N$) with  $k=10$.  This  model is
equivalent to the $M=2$ MKR in Fig. 4.  The several curves correspond, in
descending  order  at  $N=15$,  to  the following  values  of  $\tau$  in
(\ref{tb2KR}): $\tau  =10^{-6},\ 10^{-5},\ 10^{-4},\  10^{-3},\ 10^{-2},\
2\cdot  10^{-2},\ 5\cdot  10^{-2},\  0.1,\ 0.125,\  0.15,\ 0.2,\  0.175,\
1.0,\  0.25$.  Notice  the almost  coincidence  of the  curves for  $\tau
=10^{-6}$ and  $\tau =10^{-5}$, indicating  very close proximity  to QAR.
See  details  concerning  the  calculation of  $\gamma_N$  in  the  text.
\label{fig7}}
\figure{Minimal Lyapunov exponents  $\gamma_{N=12}$ and $\gamma_{N=1}$ as
a function of  $\tau$ near QAR for  the same model as in  Fig. 7.  Notice
that $\gamma_1$ appears to be  almost independent of $\tau$, assuming the
value corresponding  to QAR-localization  ($\tau \rightarrow 0$)  for all
$\tau$.\label{fig8}}
 

\begin{references}
\bibitem{not}  The term  ``dynamical localization"  is mainly  used in  a
situation  of  global chaotic  diffusion,  where  the extended  classical
motion is in sharp contrast with the localized quantum motion (see, e.g.,
Ref. \cite{qc}).  However, the localization in pseudorandom tight-binding
models    takes   place,    of   course,    also   in    local-chaos   or
almost-integrability regimes, where the  classical motion is localized as
the  quantum one.   As shown  in Refs.  \cite{fish,gri}, the  fundamental
reason for this localization is  the pseudorandom disorder, produced by a
sufficiently irrational  $\eta$.  The nature  of chaos (local  or global)
and its  strength are expected  to affect  only several features  of this
localization, such  as the localization  length.  We shall  therefore use
the term ``dynamical  localization" in this paper also  in local-chaos or
almost-integrability regimes.
\bibitem{note}  This invariance  could  be expected  from  the fact  that
$k_{eff}$ is  the parameter  in Eq. (\ref{eq}),  which determines  the QE
spectrum in  the infinitesimal  vicinity of QAR.   Since the  QE spectrum
does not  depend on  the time  limits defining the  basic period  $T$, in
particular on cyclic  permutations of the sequence  $c_j$, $k_{eff}$ must
be invariant under such permutations.
\end{references}
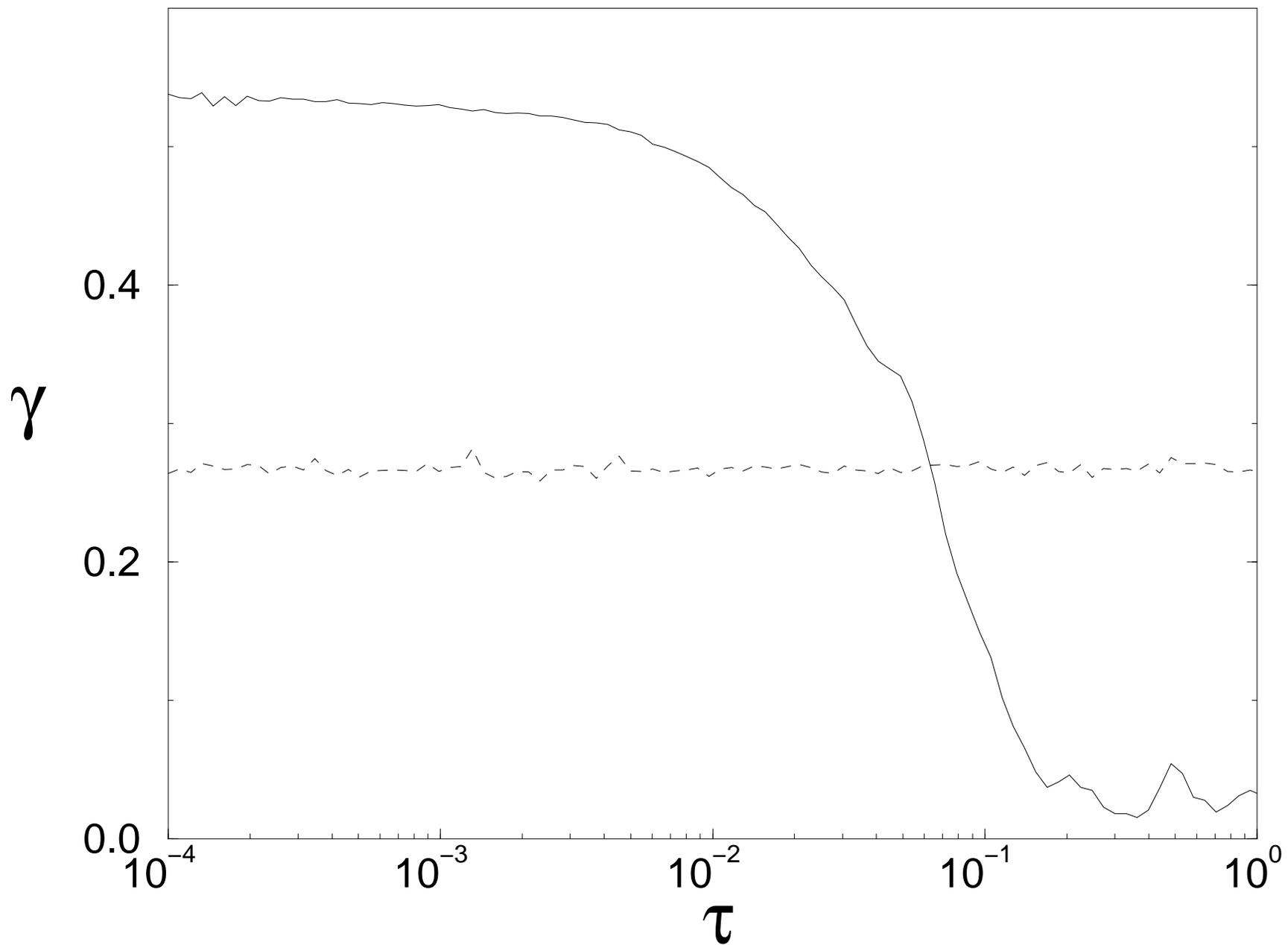
\end{document}